\documentclass[twocolumn]{aastex631}

\newcommand{\Blos}{\ensuremath{B_\mathrm{LOS}}}

\begin{document}

\title{Solar Atmospheric Heating Due to Small-scale Events in an Emerging Flux Region}

\author{Rahul Yadav}
\affiliation{Laboratory for Atmospheric and Space Physics, University of Colorado, Boulder, CO 80303, USA; \textnormal{rahul.yadav@lasp.colorado.edu}}

\author{Maria D. Kazachenko}
\affiliation{Laboratory for Atmospheric and Space Physics, University of Colorado, Boulder, CO 80303, USA; \textnormal{rahul.yadav@lasp.colorado.edu}}
\affiliation{ National Solar Observatory, 3665 Discovery Drive, 80303, Boulder, CO, USA}
\affiliation{Dept. of Astrophysical and Planetary Sciences, University of Colorado, Boulder, 2000 Colorado Ave, 80305, Boulder, CO, USA}

\author{Andrey N. Afanasyev}
\affiliation{Laboratory for Atmospheric and Space Physics, University of Colorado, Boulder, CO 80303, USA; \textnormal{rahul.yadav@lasp.colorado.edu}}
\affiliation{ National Solar Observatory, 3665 Discovery Drive, 80303, Boulder, CO, USA}
\affiliation{DKIST Ambassador}
\affiliation{Institute of Solar-Terrestrial Physics of SB RAS, Irkutsk, Russia}

\author{Jaime de la Cruz Rodr\'iguez}
\affiliation{Institute for Solar Physics, Dept. of Astronomy, Stockholm University, AlbaNova University Centre, SE-10691 Stockholm, Sweden }

\author{Jorrit Leenaarts}
\affiliation{Institute for Solar Physics, Dept. of Astronomy, Stockholm University, AlbaNova University Centre, SE-10691 Stockholm, Sweden }

\begin{abstract}
We investigate the thermal, kinematic and magnetic structure of small-scale heating events in an emerging flux region (EFR). We use high-resolution multi-line observations (including Ca II 8542~\AA, Ca II K, and Fe I 6301~\AA\ line pair) of an EFR located close to the disk center from the CRISP and CHROMIS instruments at the Swedish 1-m Solar Telescope. We perform non-LTE inversions of multiple spectral lines to infer the temperature, velocity, and magnetic field structure of the heating events. Additionally, we use the data-driven Coronal Global Evolutionary Model to simulate the evolution of the 3D magnetic field configuration above the events and understand their dynamics. Furthermore, we analyze the differential emission measure to gain insights into the heating of the coronal plasma in the EFR. Our analysis  reveals the presence of numerous small-scale heating events in the EFR, primarily located at polarity inversion lines of bipolar structures. These events not only heat the lower atmosphere but also significantly heat the corona. The data-driven simulations, along with the observed enhancement of currents and Poynting flux, suggest that magnetic reconnection in the lower atmosphere is likely responsible for the observed heating at these sites.

 \end{abstract}


\section{Introduction}
Emerging flux regions (EFRs), which are commonly found on the solar surface, are formed when flux tubes rise from the convection zone to the solar surface due to magnetic buoyancy or Parker instability \citep{1955ApJ...121..491P}. A wide range of solar activities occurs when the emerging field lines rise and pass through different layers of the solar atmosphere \citep{1993ASPC...46..471C,2014LRSP...11....3C}. Therefore, EFRs play an important role in understanding the interplay between different layers of the solar atmosphere.

The magnetic field can emerge anywhere on the solar surface at various spatial scales \citep{2009ApJ...698...75P, 2011PASJ...63.1047O}. Typically, during the emergence of field lines, two main patches with opposite polarities move apart from each other, and multiple small-scale magnetic bipolar regions appear between them in the photosphere. Moreover, the magnetic field associated with these bipolar regions rises and interacts with pre-existing magnetic field in the chromosphere or corona, leading to magnetic reconnection. This reconnection process gives rise to various heating events such as Ellerman bombs, UV bursts, transient brightenings, chromospheric jets, or flares. Such scenario of EFR has been seen in various observations \citep{2002ApJ...574.1074S,2014Sci...346C.315P,2015ApJ...812...11V,2017ApJS..229....4C,2017ApJ...836...63T, 2018ApJ...856..127G, 2018A&A...612A..28L,2019ApJ...887...56T,2019A&A...632A.112Y,2020A&A...633A..58O, 2022ApJ...933...12M,2022ApJ...929..103T, 2023A&A...673A..11R} and numerical simulations \citep{2008ApJ...687.1373C, 2017A&A...601A.122D, 2019A&A...626A..33H}.

To understand small-scale heating events, several mechanisms have been proposed (e.g., MHD wave and magnetic heating, see \citealt{1990SSRv...54..377N, 2018ApJ...862L..24P} and references therein). However, the process by which energy is transported from the photosphere to the higher layers during the emergence of small-scale bipolar regions in EFRs remains unclear \citep{1977ARA&A..15..363W, 1990SSRv...54..377N}. Recently, \citealt{2018ApJ...862L..24P} proposed a theoretical model for chromospheric and coronal heating by considering a bipolar converging region. They demonstrated that two opposite-polarity regions having equal magnetic flux, situated below horizontal magnetic field, will undergo magnetic reconnection driven by flux cancellation if their separation is smaller than the flux interaction distance \citep{1998ApJ...507..433L}. The energy released during the magnetic reconnection can then heat the chromosphere and corona located above.

The magnetic field plays a crucial role in explaining the observed heating events in different layers. 
Recent observations have also shown that magnetic flux cancellation in the photosphere, both in quiet-Sun regions and EFRs, is associated with intense brightening observed in the chromosphere and corona \citep{2018ApJ...857...48G, 2018A&A...612A..28L, 2019ApJ...887...56T,2021A&A...647A.188D,2021ApJ...909..133M,2021ApJ...921L..20P, 2023MNRAS.521.3882K}. Although extensive studies have been conducted on the bipolar regions within EFRs in the photosphere, utilizing data from various ground-and-space based telescopes, simultaneous investigations involving photospheric and chromospheric vector magnetograms are still rare due to limited chromospheric observations.
Recently, utilizing multi-line spectropolarimetric observations of an EFR, \citealt{2018A&A...612A..28L}, demonstrated a correlation between the \ion{Ca}{2}~K intensity and the horizontal field strength in the chromosphere. Furthermore, for the same EFR, \citealt{2021A&A...647A.188D} found that the radiative losses in the chromosphere can reach up to 160~kW~m$^{-2}$ at the reconnection site.

In this study, we present an analysis of small-scale bipolar regions within EFR to understand their thermal, kinematic, and magnetic structure. We also investigate their impact on the chromospheric and coronal heating. To achieve this, we employ a combination of multi-line spectropolarimetric observations in the photosphere and chromosphere, co-aligned coronal images, and data-driven simulations to elucidate the field topology above the bipolar regions. Non-local thermodynamic equilibrium inversions of multiple spectral lines, including the \ion{Fe}{1}~6301 Å line pair, \ion{Ca}{2}~8542 Å, and \ion{Ca}{2}~K, are utilized to derive the stratification of physical parameters such as temperature, line-of-sight velocity, magnetic field, and micro-turbulent velocity. These derived parameters are then employed to investigate the signatures of magnetic reconnection at different heights. Additionally, we estimate the physical quantities described in the heating model proposed by \citealt{2018ApJ...862L..24P}.

Section \ref{Sec:observation} describe our observations. Sections \ref{sec:method_and_data}
 and \ref{sec:results} present our analysis and results. The obtained results are discussed in Section \ref{sec:Discussion}, and summarized in Section~\ref{sec:conclusion}.
 
\section{Observations}
\label{Sec:observation}
\begin{figure*}
    \centering
    \includegraphics[width=0.67\linewidth]{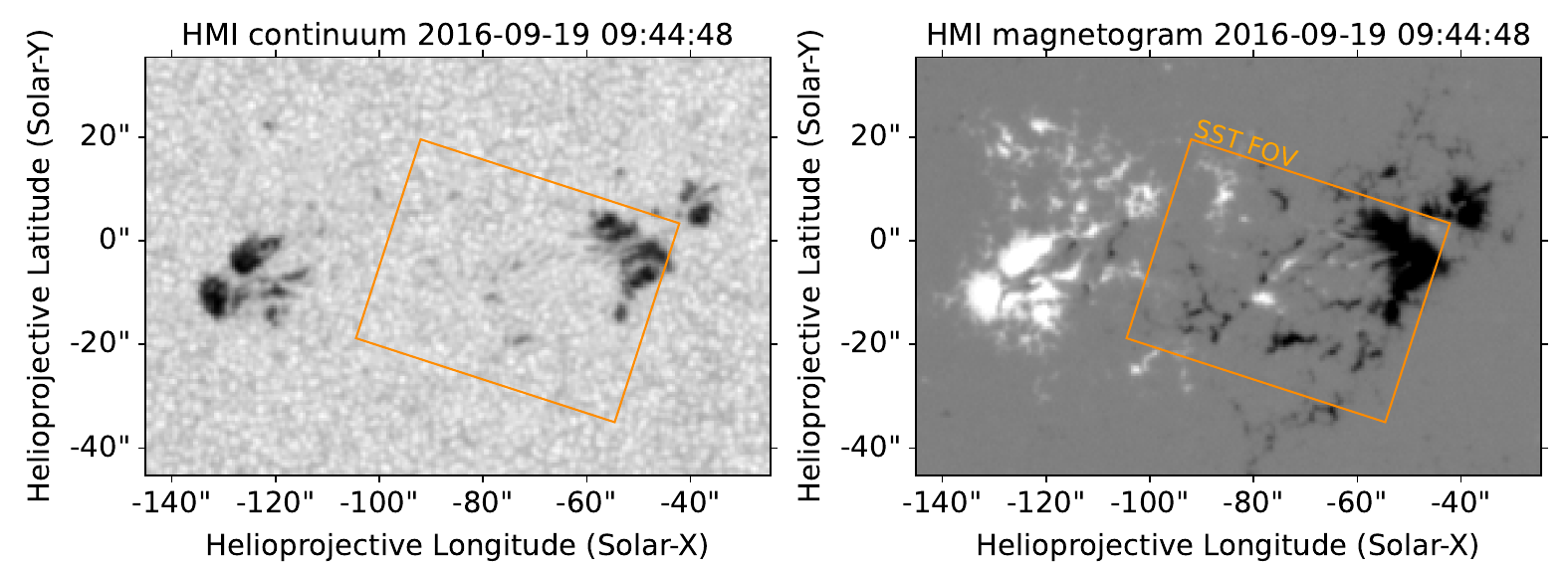}
    \includegraphics[width=0.32\linewidth]{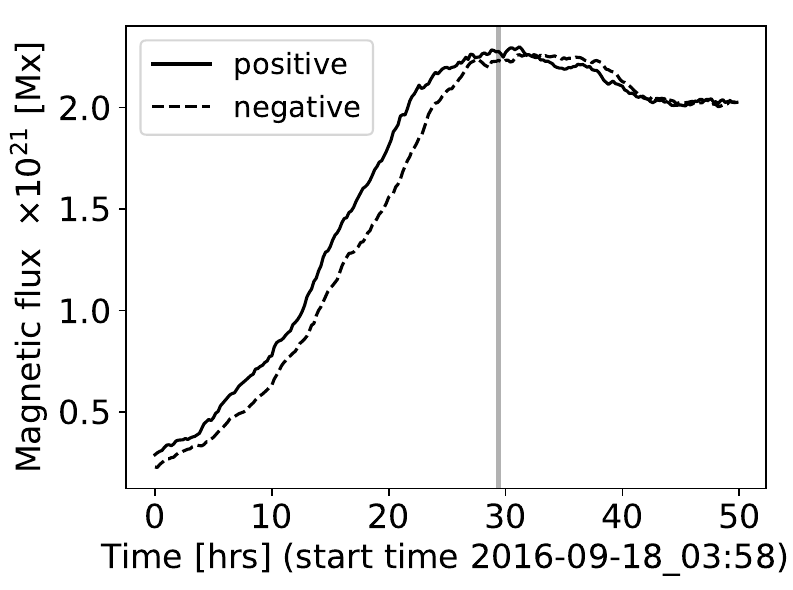}
    \caption{Overview of emerging flux region (EFR). \textit{Left and middle panels}: HMI continuum image (left) and LOS magnetogram (middle) saturated at $\pm$1000~G, where black and white colors indicate negative and positive polarities, respectively. The orange box outlines the FOV of SST. \textit{Right panel:} The temporal evolution of positive and negative magnetic fluxes in the EFR are indicated by solid and dashed black line, respectively. The vertical gray line marks the time of the SST observation.}
    \label{fig:sstfov}
\end{figure*}

\subsection{Target and Data Reduction}
We use observations of the active region (AR) NOAA 12593, located close to disk center ($\mu$=1.0), recorded between 09:31 and 09:57 UT
on September 19, 2016, with the CRisp Imaging SpectroPolarimeter
(CRISP; \citealt{Scharmer2008}) and the CHROMospheric Imaging Spectrometer (CHROMIS; \citealt{2017psio.confE..85S}) instruments at the Swedish Solar Telescope (SST; \citealt{2003SPIE.4853..341S}). 

The CRISP simultaneously recorded full spectropolarimetric data in the \ion{Ca}{2} 8542 \AA~and \ion{Fe}{1}~6301~\AA~line pair. The \ion{Ca}{2} 8542 \AA~line scans consisted of 21 wavelength positions spanning a range of 1.7~\AA\ around line center, with steps of 0.765~\AA~in the inner wings and two wavelength positions at $\pm$1.7~\AA~relative to line center. The \ion{Fe}{1}~6301~\AA~spectral line was scanned with nine equidistant wavelength positions spanning a range of 0.19~\AA\ around line center, whereas the \ion{Fe}{1}~6302 ~\AA\ was scanned with seven equidistant wavelength positions spanning a range of 0.28~\AA\ around line center. The CRISP data were obtained with a cadence of 36.6~s.

The CHROMIS recorded \ion{Ca}{2}~K intensity profiles at 39 wavelength positions spanning a range of 1.33~\AA\ around line center, with 37 evenly spaced steps of 0.038\AA\ in the inner wings and two wavelength positions at $\pm{1.33}$~\AA\ relative to the line center. In addition to this, one point in the continuum at 4000~\AA\ was also observed with the CHROMIS instrument. 

The CRISP data were reduced using the CRISPRED \citep{2015A&A...573A..40D} post-processing pipeline, which includes image reconstruction through multi-object multi-frame blind deconvolution (MOMFBD; \citealt{2005SoPh..228..191V}) and removal of small-scale seeing-induced deformations. The CHROMIS data were reduced using the CHROMISRED pipeline \citep{2021A&A...653A..68L}. The CRISP data were aligned with the CHROMIS data and resampled to the CHROMIS pixel scale of 0.0375\arcsec. As the CRISP data were obtained with a lower cadence, we interpolated the CRISP data to the CHROMIS cadence using nearest-neighbor interpolation. For all data, the intensity calibration was performed with the quiet Sun data located close to the disk center after taking into account the limb darkening, whereas the absolute wavelengths were calibrated with the atlas profiles given by \cite{1984SoPh...90..205N}. This AR is also analyzed by \citealt{2018A&A...612A..28L} and \citealt{2021A&A...647A.188D}.

During our analysis, we also utilized ultraviolet (UV) and extreme ultraviolet (EUV) images observed by the Atmosphere Imaging Assembly (AIA; \citealt{2012SoPh..275...17L}), as well as full-disk continuum images and vector magnetograms from the Helioseismic and Magnetic Imager (HMI; \citealt{2012SoPh..275..207S}) aboard the Solar Dynamic Observatory (SDO; \cite{2012SoPh..275....3P}). The AIA takes full-disk images in seven EUV bands with a cadence of 12~sec and in two UV bands at 1600~\AA\ and 1700~\AA\ with a cadence of 24~sec. The spatial scale of AIA images is 0.6\arcsec per pixel. All AIA and HMI images were corrected using the standard solar software (SSW) routines (e.g., \texttt{aia\_prep.pro} and \texttt{hmi\_prep.pro}). Finally, all AIA, HMI, CRISP, and CHROMIS data were co-aligned through image cross-correlation. 

\subsection{Overview of Observations}

\begin{figure*}
    \centering
    \includegraphics[width=1\linewidth]{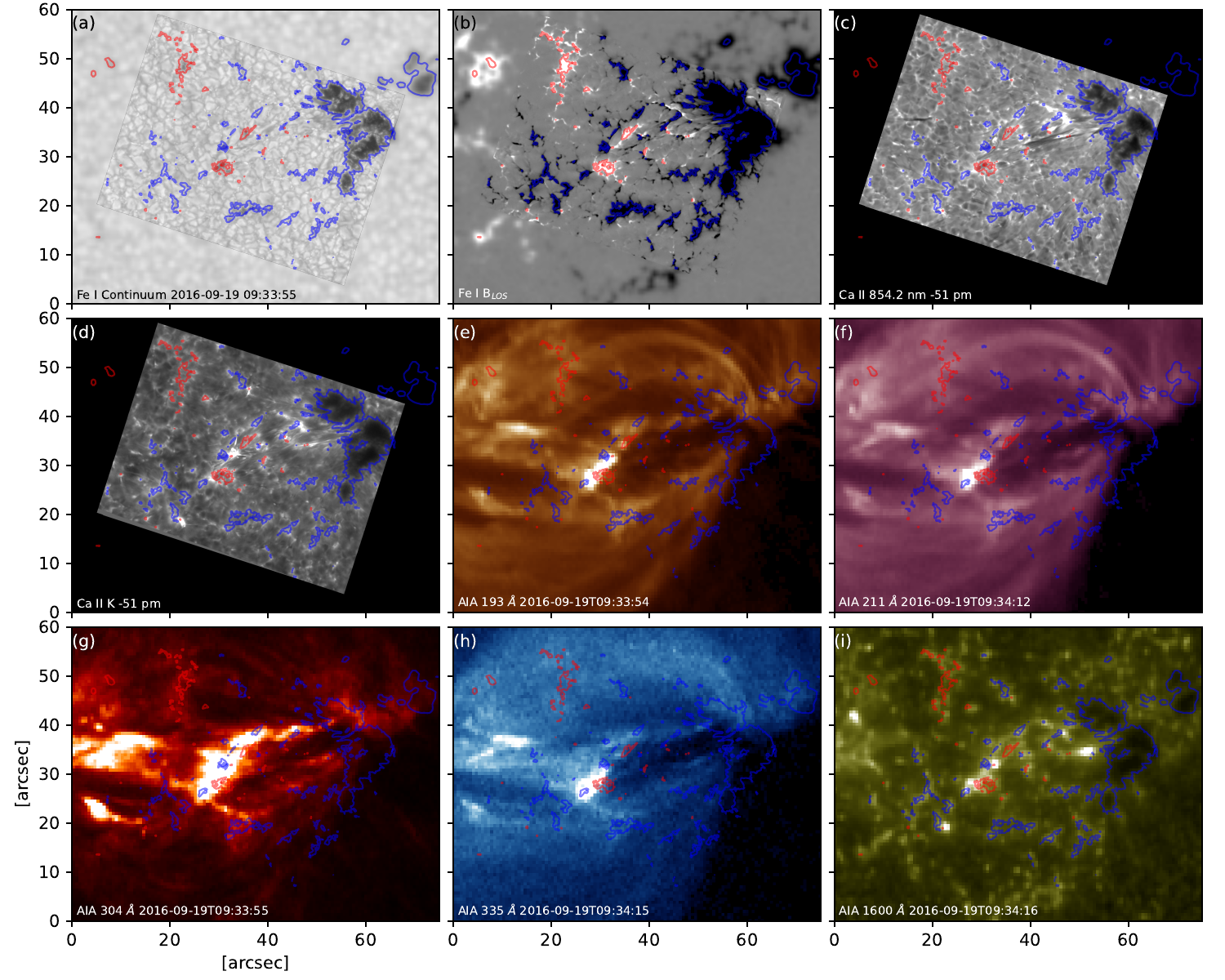}
    \caption{Overview of SST and SDO observations taken on 2016-09-19. \textit{Panel a}: continuum intensity at \ion{Fe}{1}~6302~\AA~superimposed on the HMI continuum intensity; \textit{Panel b}: Line-of-sight magnetic field inferred from the Milne-Eddington inversion of the \ion{Fe}{1}~6301 line pair superimposed over HMI magnetogram; \textit{Panels c and d}: chromospheric intensity maps, observed with the
CRISP and CHROMIS instruments at the SST; \textit{Panels e--i}: AIA images observed in different channels. The blue and red contours indicates negative and positive polarity at the level of 800~G in the photosphere. Solar north is up.}  
    \label{fig:overview}
\end{figure*}

Figure~\ref{fig:sstfov} shows an overview of the AR 12593 that started emerging on September 18, 2016 around 04:00 UT. In order to study the history of the AR, we calculated the magnetic flux evolution in the FOV using the magnetic parameters obtained from the Space-weather HMI Active Region Patch (SHARP,  \citealt{2014SoPh..289.3549B}). The calculated magnetic flux for positive and negative polarities, using the B$_z$ component of the magnetic field strength, are depicted in Fig.~\ref{fig:sstfov}. During the flux emergence period, the flux of either polarity has increased up to $\sim$10$^{21}$~Mx. Furthermore, the peak flux emergence rate for the AR is $2.2\times$10$^{17}$~Mx~s$^{-1}$. As shown in the Fig.~\ref{fig:sstfov} the SST observations were performed as the emergence of AR 12593 was ending. 

The SST recorded FOV closer to the negative polarity of the EFR including regions having bipolar structures as shown in Fig.~\ref{fig:overview}, that we refer to as mixed polarity regions. During our observations, brightening events were noticed in different SDO/AIA filtergrams mainly close to these mixed polarity regions highlighted by blue and red contours. The intense brightening in \ion{Ca}{2}~8542, \ion{Ca}{2}~K and AIA images also indicates that the region located above the mixed polarity region is heated significantly.

\begin{figure*}[!ht]
    \centering
    \includegraphics[clip,trim=0.cm  0.1cm 0.cm 0.cm,width=1\textwidth]{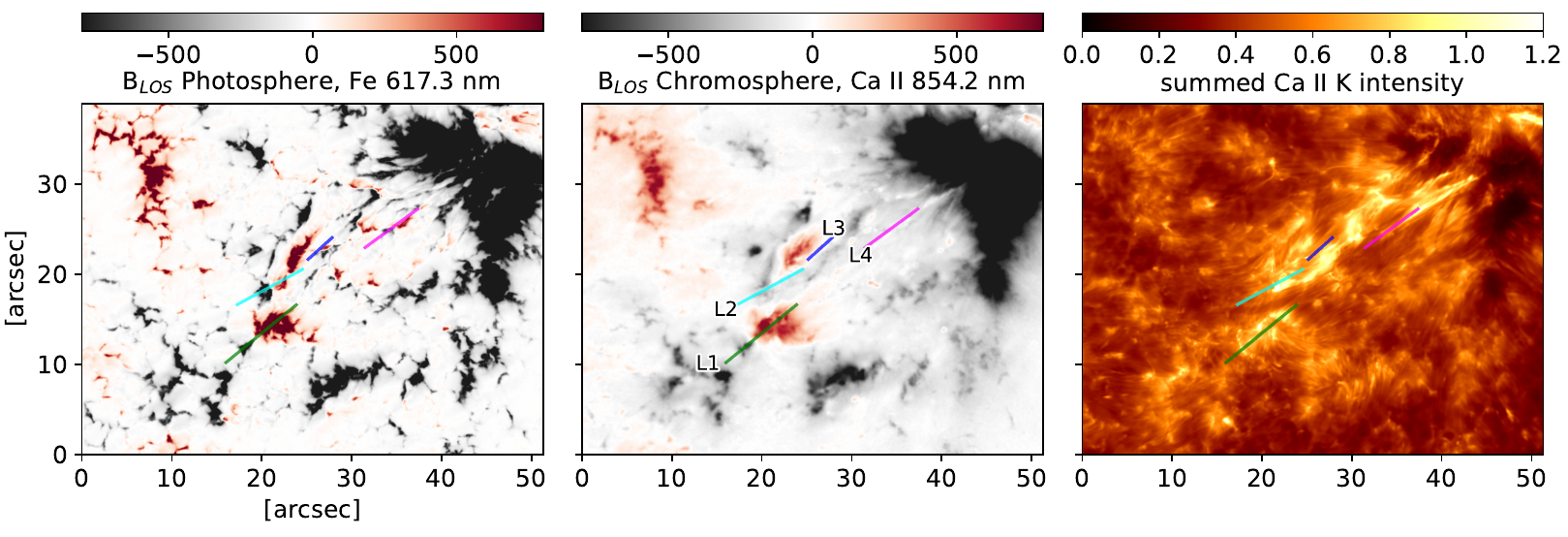}
    
    \includegraphics[clip,trim=0cm  0.5cm 0.cm .4cm, width=1\textwidth]{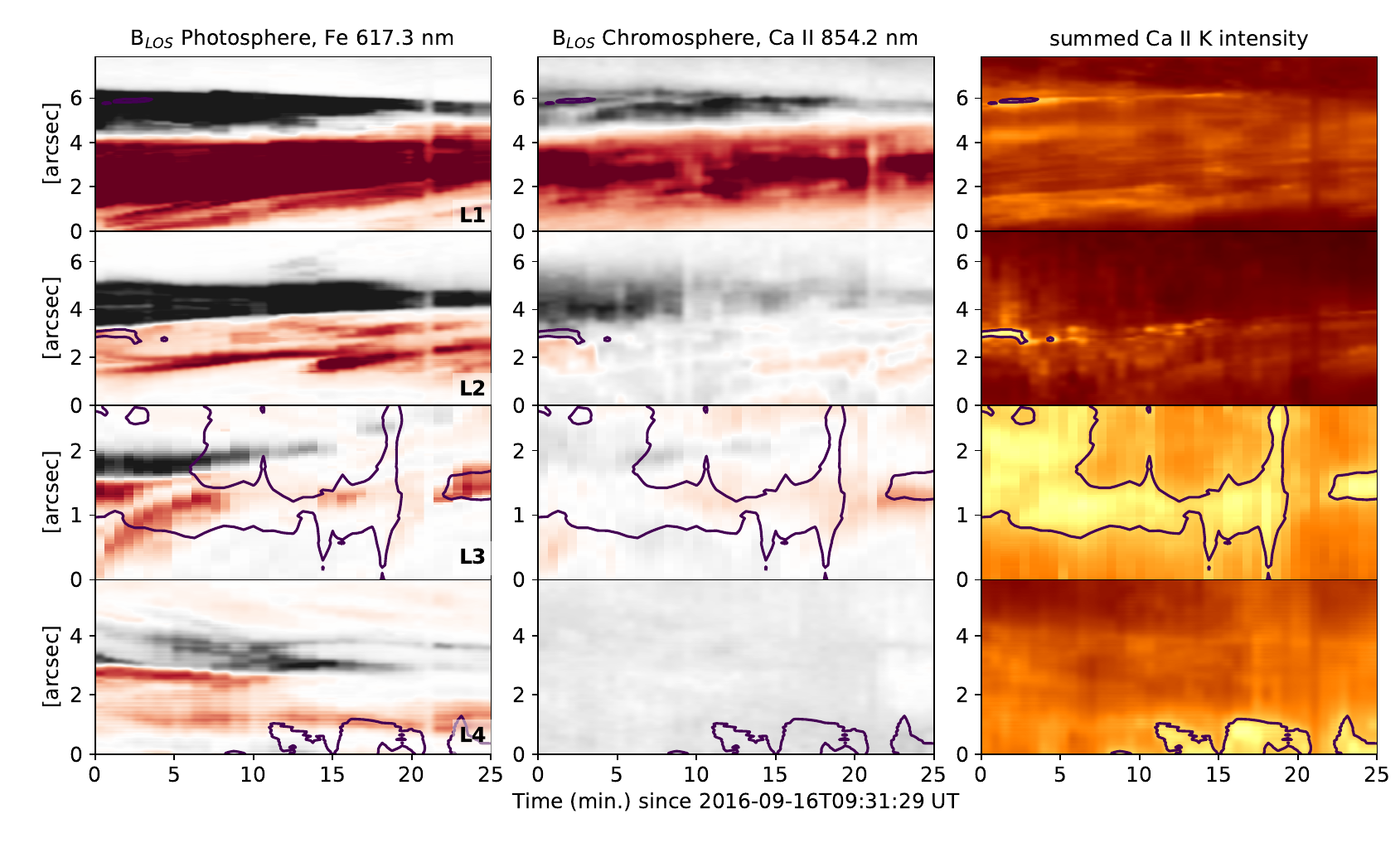}
    \caption{Time and slit plots of magnetic field and \ion{Ca}{2}~K intensity. \textit{Top panel:} The photospheric (left) and chromospheric (middle) \Blos, and \ion{Ca}{2}~K wavelength summed intensity (right). \textit{Three bottom panels}: Temporal evolution of photospheric and chromospheric $B_\mathrm{LOS}$, and \ion{Ca}{2}~K wavelength summed intensity across four slits highlighted in the top panels. In each panels the black contours refer to the enhanced intensity in the \ion{Ca}{2}~K line.}
    \label{fig:blos_photo_chromo_slice}
\end{figure*}

\begin{figure*}[!ht]
    \centering
        \includegraphics[width=0.91\linewidth]{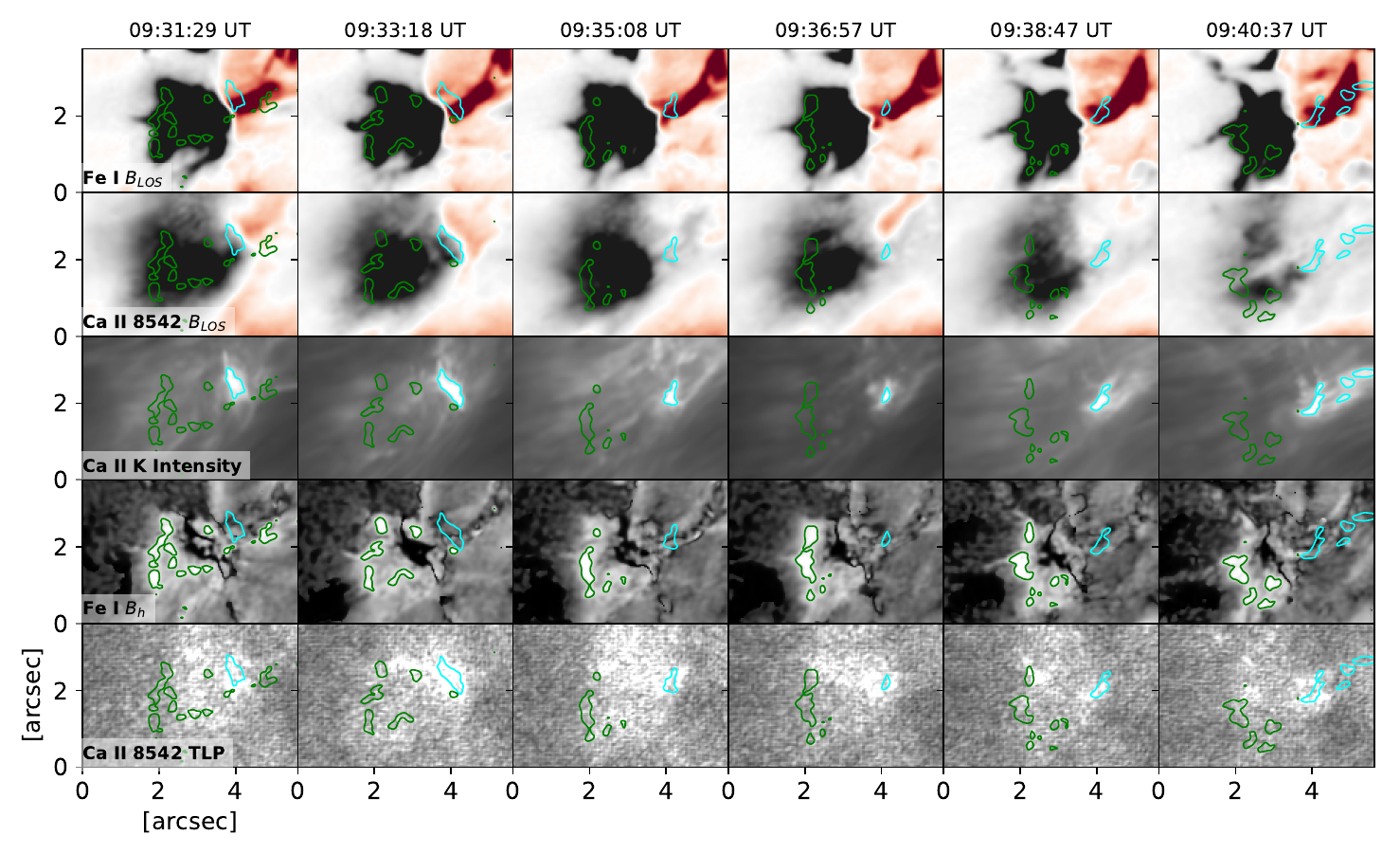}
       \hspace{-3mm} \includegraphics[clip,trim=0.1cm  -1.7cm 0.cm -0.cm,width=0.06\linewidth]{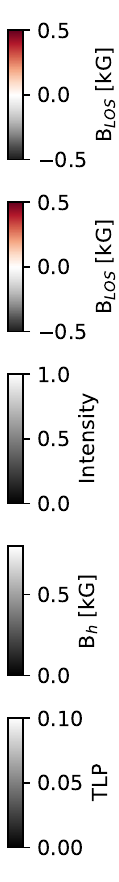}
    \caption{Temporal evolution of a bipolar region across the L2 slit shown in Fig.~\ref{fig:blos_photo_chromo_slice}. The top two panels show the B$_{LOS}$ in the photosphere and the chromosphere. Middle panel shows the wavelength summed intensity of \ion{Ca}{2}~K line. The bottom two panels show the horizontal magnetic field ($B_h$) in the photosphere and total linear polarization (TLP) in the chromosphere. The green and cyan contours indicate the locations of strong (700~G) $B_h$ (\ion{Fe}{1}~6173~\AA) and \ion{Ca}{2}~K intensity, respectively.}
    \label{fig:bipolar1}
\end{figure*}

\begin{figure*}
    \centering
    \includegraphics[width=0.92\linewidth]{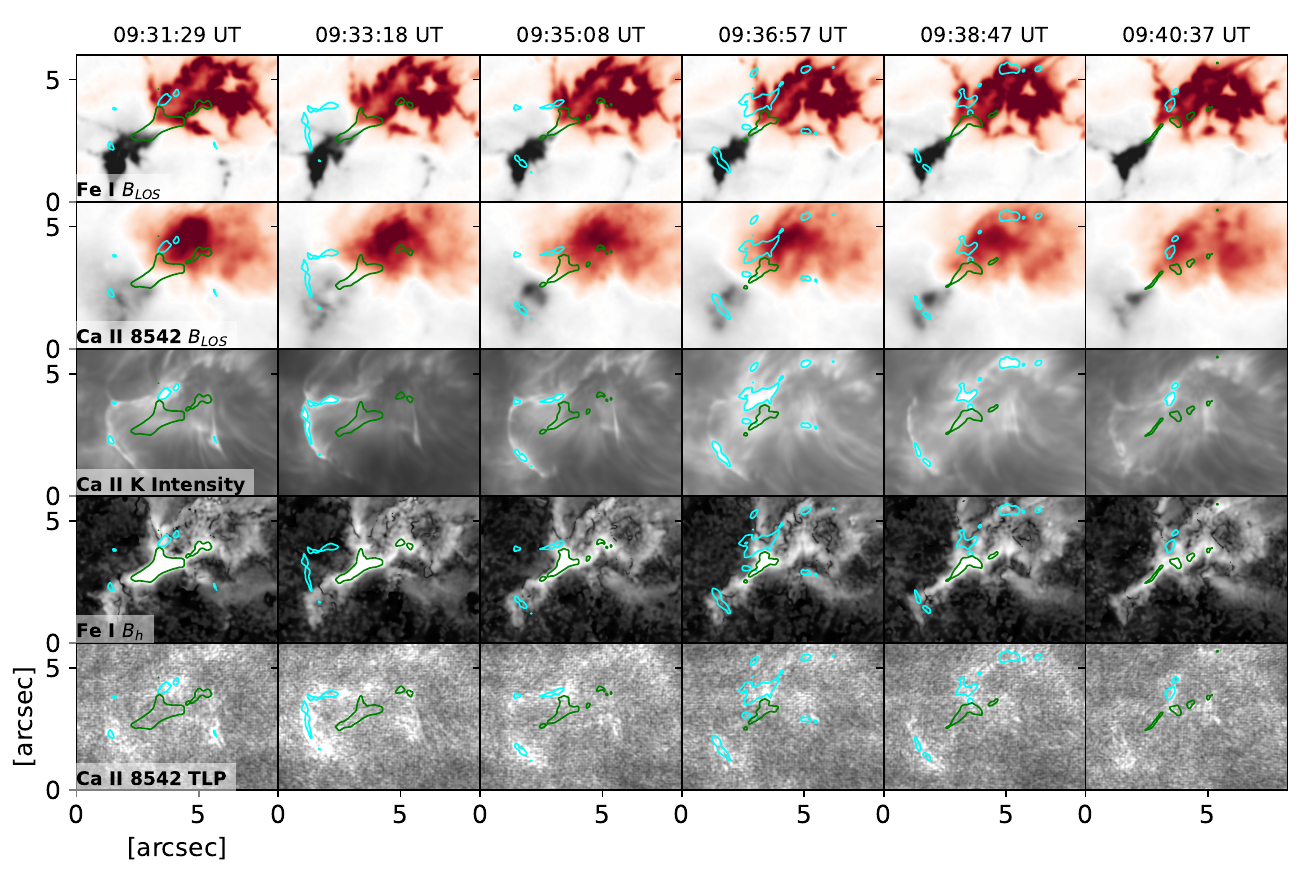}
        \hspace{-2mm}\includegraphics[clip,trim=0.25cm  -1.5cm -0.1cm -0.1cm,width=0.075\linewidth]{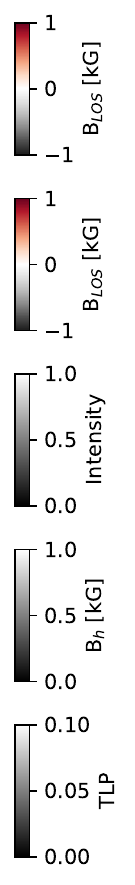}
    \caption{Same as Figure~\ref{fig:bipolar1}, but for a region across the L1 slit shown in Fig. \ref{fig:blos_photo_chromo_slice}.}
    \label{fig:bipolar2}
\end{figure*}

\section{Methods and Data Analysis}\label{sec:method_and_data}
\subsection{Inversion of the Spectropolarimetric Data}
The physical parameters such as magnetic field vector and line-of-sight (LOS) velocity were inferred by inverting the photospheric spectral line \ion{Fe}{1}~6301~\AA\ using a Milne-Eddington SPIN code 
\citep{2017SoPh..292..105Y}. 
Then we resolved the 180$^{\circ}$ azimuthal ambiguity using the automated ambiguity resolution code \citep{2014ascl.soft04007L}, which is based on the minimum energy method \citep{1994SoPh..155..235M}. Furthermore, we employed a spatially regularized Weak Field Approximation (WFA; \citealt{2020A&A...642A.210M}) method to infer the LOS magnetic field from the \ion{Ca}{2}~8542 observations. The linear polarization signal was not sufficient to infer the magnetic field vector in the chromosphere using the \ion{Ca}{2}~8542 line.

To estimate the stratification of the physical parameters such as temperature, magnetic field, LOS velocity, and microturbulent velocity, we inverted the Stokes profiles of \ion{Fe}{1}~6301 line pair, \ion{Ca}{2}~8542 and \ion{Ca}{2}~K line simultaneously using the multi-line inversion STiC code 
\cite{Jaime2019}.
 We inverted all four Stokes parameters in the \ion{Fe}{1}~6173~\AA\ and \ion{Ca}{2}~8542~\AA\ lines, but only Stokes $I$ in the \ion{Ca}{2}~K line. The STiC inversion code is built around a modified version of the RH code \citep{Uitenbroek2001} in order to derive the atomic populations by assuming statistical equilibrium and a plane-parallel geometry. The equation of state is borrowed from the Spectroscopy Made Easy (SME) computer code described in \cite{2017A&A...597A..16P}. The radiative transport equation is solved using cubic Bezier solvers \citep{2013ApJ...764...33D}. During inversion, we considered the \ion{Ca}{2}~8542~\AA\ line in non-LTE conditions, under the assumption of complete frequency redistribution, while the \ion{Ca}{2}~K line was synthesized in non-LTE conditions with partial redistribution effects of scattered photons 
 \citep{2012A&A...543A.109L}.
\subsection{Differential Emission Measure}

We perform the Differential Emission Measure (DEM) analysis of AIA/SDO data to investigate the temperature distribution of the plasma. The DEM analysis involves solving an inverse problem: inferring the temperature distribution of the plasma from the observed intensities. 
The measured intensity ($I_\lambda$) for a given AIA channel can be expressed as 
\begin{equation}
    I_\lambda = \int K_\lambda(T) DEM(T) dT,
\end{equation}
where $K_\lambda(T)$ refers to the response function of the corresponding AIA channel. We utilized the regularized inversion code developed by \citealt{2012A&A...539A.146H} to derive the DEM maps from the aligned AIA channels.
We employed six EUV channels (94, 131, 171, 193, 211, and 355 Å) of the AIA instrument aboard the SDO. These specific wavelength channels are sensitive to emissions originating from different ionization states of various elements, providing a wide temperature coverage. 

\subsection{Data-driven Simulation of the AR 12593}
\label{sec:cgem_intro}
To understand the evolution of the 3D magnetic field configuration above the AR, we performed a data-driven simulation using the Coronal Global Evolutionary Model (CGEM; \citealt{2020ApJS..250...28H}). The CGEM uses a time sequence of electric field maps derived from photospheric vector magnetograms and Dopplergrams, to derive a time-dependent, magnetogrictional nonpotential model for the magnetic field. The photospheric magnetic fields and Doppler elocity are obtained from the HMI/SDO, whereas the electric field patches are computed using the ``PDFI" inversion method \citep{2014ApJ...795...17K,2020ApJS..248....2F} in the photosphere. For the vector magnetogram  we utilized the JSOC SHARP number 6764 of AR 12593. We perform simulations of the EFR from September 19, 2016 (06:00 UT) to September 20 2016 (11:00 UT), covering the full time domain of SST observations. The simulations are performed in a 3D domain of 672$\times$372$\times$336 grid points with a grid spacing of 0\arcsec.5, which is similar to the spatial resolution of SDO/HMI. We set the simulation output time step to 24~seconds, which is less than the cadence of SST/CRISP spectropolarimetric observations. The spatial and temporal cadence in the simulation is enough to investigate the magnetic field topology of the bipolar regions, which generally have size of more than an arcsec in our observations.

\section{Results}
\label{sec:results}
\subsection{Mixed Polarity Region in the Photosphere and the Chromosphere}
From the high-resolution observations of the EFR, in Figure~\ref{fig:blos_photo_chromo_slice} several small-scale mixed-polarity regions can be clearly identified both in the photosphere and the chromosphere. Strong patches of the line-of-sight magnetic field (\Blos) located in the photosphere can also be noticed in the chromosphere, but with a reduced strength of the magnetic field.
Additionally, $\Blos$ in the chromosphere covers a larger area compared to the photosphere, where they are located in small and compact regions.

Near the mixed-polarity region, we also observed a significant intensity enhancement in the \ion{Ca}{2} lines, as shown in Figure~\ref{fig:blos_photo_chromo_slice}. Additionally, as demonstrated in Fig.\ref{fig:overview}, we observe intensity enhancement in all channels of AIA. Such enhancement suggests that a significant amount of energy is released due to magnetic flux cancellation leading to magnetic reconnection in the lower atmosphere, which can heat plasma at different heights. 

In Figure \ref{fig:blos_photo_chromo_slice}, $B_\mathrm{LOS}$ in the photosphere and the chromosphere are derived using the ME inversion and the WFA, respectively. 
The time-distance diagram taken across the selected slits, passing through the mixed polarity regions, showed that in certain instances (L2 and L3), the location of increased intensity (in the \ion{Ca}{2}~K line) was situated over the polarity inversion line (PIL) of the mixed polarity regions. At this location, $\Blos$ decreased in both the photosphere and the chromosphere. However, in other cases (L1 and L4), intensity enhancement was not observed directly above the PIL, but was instead observed slightly away from it. We note that the slit width is of pixel-size and may not capture the brightening location at all position. 

The Stokes Q and U signals in the \ion{Ca}{2}~8542~\AA\ are weak throughout the FOV, except the locations of pores. To estimate a proxy for linear polarization in the chromosphere, we generated the total linear polarization (TLP) maps, using a methodology similar to that employed by \cite{2018A&A...612A..28L}. The TLP maps are computed as $\sum_{i=0}^{n}\sqrt{Q^2_i+ U^2_i},$
where the summation of Stokes Q and U profiles is performed over all wavelength positions. Such TLP maps provide qualitative information about the horizontal component of magnetic field, a stronger TLP means stronger horizontal magnetic field. 

Figure \ref{fig:bipolar1} demonstrates the temporal evolution of a mixed polarity region covering the L2 slit highlighted in Figure \ref{fig:blos_photo_chromo_slice}. The maps illustrate that the intensity enhancement in the \ion{Ca}{2}~K line lies on the PIL. Although the brightness varies with time, it remains near or on the PIL throughout the time domain of the SST. Additionally, magnetic flux cancellation is observed in both the photospheric and chromospheric LOS magnetic fields. 
In the chromosphere, high TLP observed at the location of \ion{Ca}{2}~K brightening, which is also reported by \cite{2018A&A...612A..28L}. 

 For the second bipolar region (covering L1 slit), we find significant horizontal magnetic field ($B_h$) near the PIL in the photosphere (see Fig. \ref{fig:bipolar2}). This observation indicates that the magnetic field lines tend to become more horizontal in the vicinity of the PIL. The brightening in \ion{Ca}{2}~K intensity reveals a loop-like structure connecting regions of opposite polarity. Additionally, the temporal evolution analysis of this region shows a reduction in $B_h$ within the photosphere. However, there is no corresponding change in the TLP of the chromosphere. Similar to Figure~\ref{fig:bipolar1}, strong patches of TLP in the chromosphere noticed at the \ion{Ca}{2}~K brightening sites. This observation suggests that small-scale loops located at the PIL may be disappearing due to magnetic flux cancellation, resulting in the submergence of reconnected smaller loops.

\subsection{Differential Emission Measure Analysis Over Bipolar regions}

\begin{figure*}[!ht]
    \centering
    \includegraphics[width=0.9\linewidth]{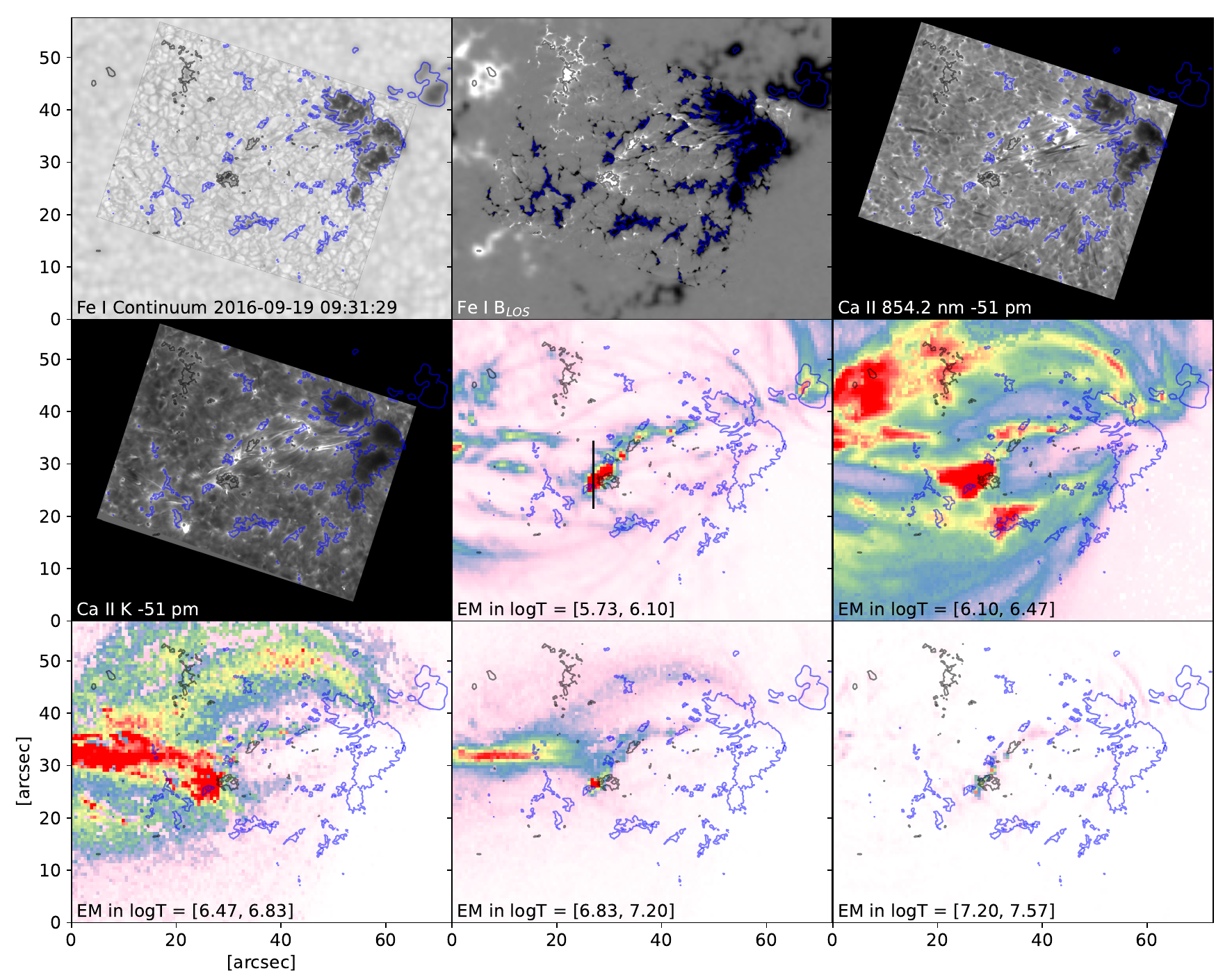}
    \includegraphics[clip,trim=0.cm  -4cm -0.cm -0.0cm, width=0.09\linewidth]{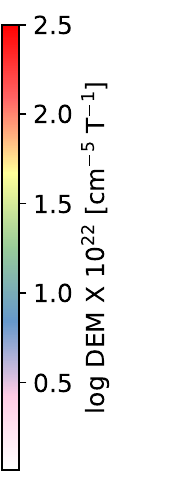}
    \caption{DEM integrated over five different temperature bins ranging from log$_{10}$ 5.7 to 7.57 [K].  First four panels showing the continuum intensity, \Blos~in the photosphere, \ion{Ca}{2}~8542 and \ion{Ca}{2}~K intensity maps. The blue and black contours indicate negative and positive polarities at the 800~G level in the photosphere.}
    \label{fig:dem_overview}
\end{figure*}

\begin{figure}[!h]
    \centering
    \includegraphics[width=0.87\linewidth]{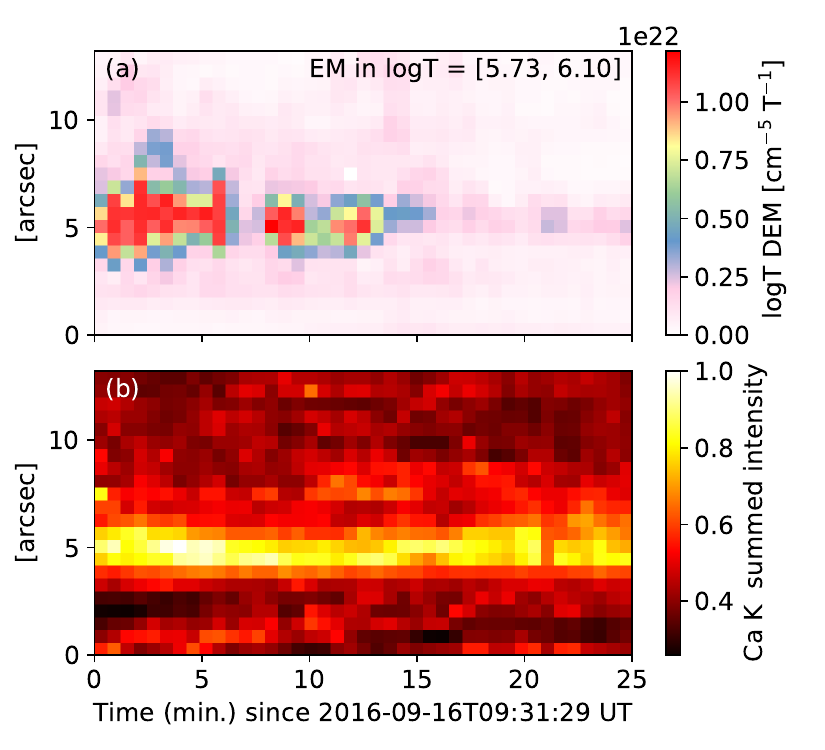}
    \caption{Temporal evolution of DEM (a) and \ion{Ca}{2}~K summed intensity (b) across a black vertical line shown in Figure~\ref{fig:dem_overview}.} 
    \label{fig:demslit}
\end{figure}

For the DEM analysis, we considered a temperature range spanning from log$_{10}$~T = 5.7 to 7.6 [K]. Figure \ref{fig:dem_overview} illustrates the integrated DEM in various temperature bins, revealing a noticeable increase in temperature (above $1$~MK) near a bipolar region. The electron density, $n_e$, is calculated as $\sqrt{EM/l}$, where the $EM$ is the integrated DEM over a temperature range of log$_{10}$~T = 6.1 to 6.5, and $l$ is the LOS length scale of the emission. We have adopted the value of $l$ as 1~Mm, which is also assumed by \citealt{2020ApJ...897...49P} for a converging bipolar region. This choice leads to an electron density of $\sim$ 0.5$\times$10$^9$~cm$^{-3}$, which is typically observed in active regions \citep{2001ApJ...550..475A}. Then we estimated the thermal energy flux from this region as $E_{th} = 3n_ek_bTl^3$, that turn out to be 4.5$\times$10$^{23}$ ergs~s$^{-1}$ (4.5$\times$10$^{7}$ ergs cm$^{-2}$ s$^{-1}$) with T=2.2 MK, which is sufficient to power the local corona \citep{1977ARA&A..15..363W}.

The temporal evolution of DEM and \ion{Ca}{2}~K intensity across a black vertical slit indicated in Figure~\ref{fig:dem_overview} is illustrated in Figure \ref{fig:demslit}. The figure shows co-temporal and co-spatial enhancements in both the DEM and \ion{Ca}{2}~K intensity. Brightening in \ion{Ca}{2}~K persists throughout the SST time domain, but the DEM does not show significant emission after $\sim$ 12 minutes. Furthermore, the figure demonstrates that heating persists for a longer duration in the chromosphere compared to the corona along the selected slit.
This observation suggests that the energy released during flux cancellation results in temperature enhancement both in the chromosphere and in the corona.

\subsection{Stratification of Physical Parameters using non-LTE Inversion.}
\begin{figure*}
    \centering
    \includegraphics[width=1\linewidth]{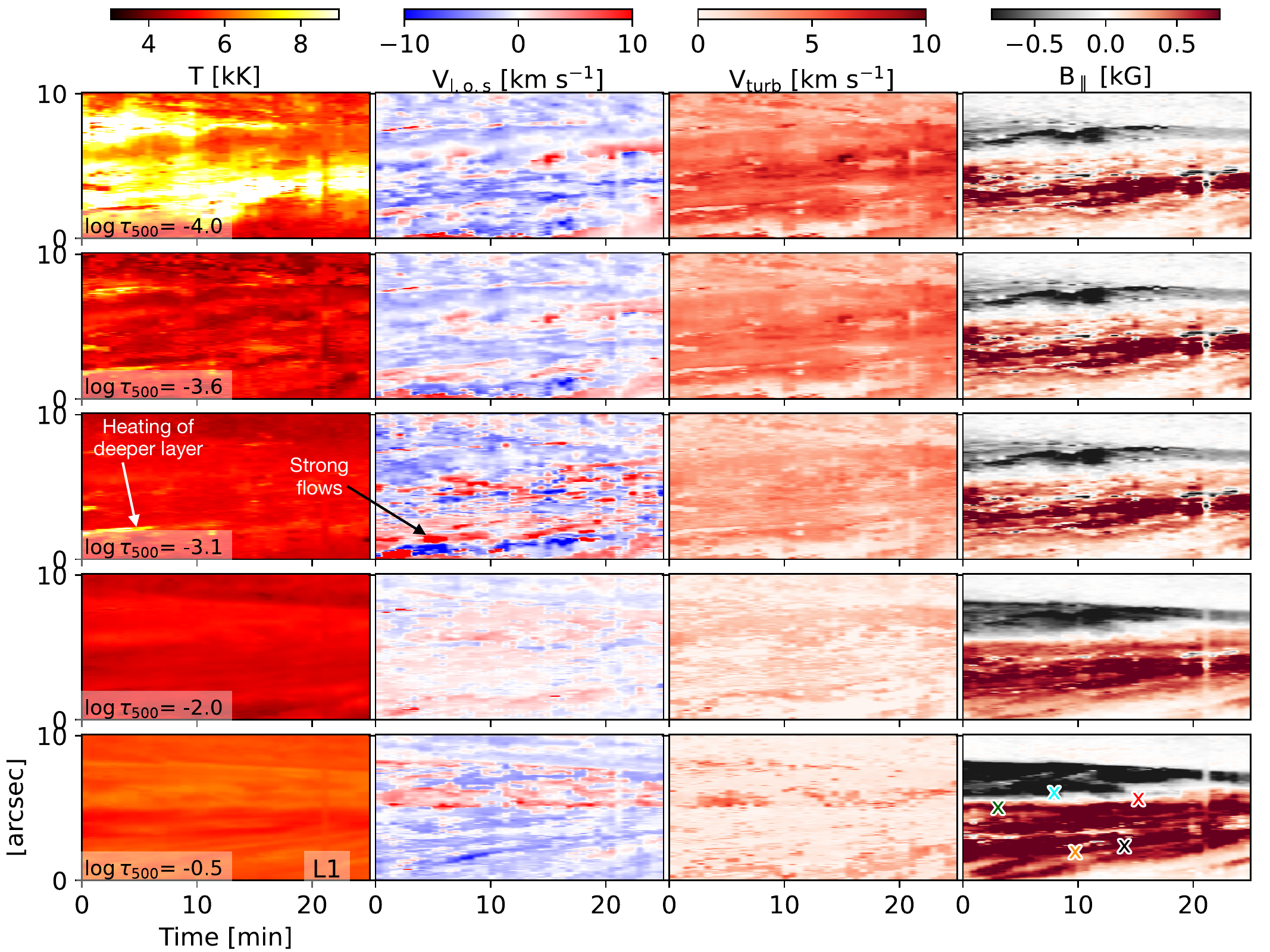}
    \caption{Temporal evolution (since 2016-09-16T09:31:29 UT) and stratification of physical parameters obtained from non-LTE inversion. For L1 slit, highlighted in Figure~\ref{fig:blos_photo_chromo_slice}, the stratification of temperature, LOS velocity, microturbulent velocity and LOS magnetic field is demonstrated at selected optical depths. The observed and best-fitted Stokes profiles at the location indicated by colored cross symbols are shown in the appendix (Figures~\ref{fig:example_fittingL1}).}
    \label{fig:invertedmaps_L1}
\end{figure*}

\begin{table}
\caption{Number of nodes used for the temperature, LOS velocity (V$_{\rm LOS}$), turbulent velocity (V$_{\rm turb}$), LOS magnetic field (B$_{\parallel}$), horizontal magnetic field (B$_{\perp}$), and azimuth ($\phi$) during each cycle of the inversion.}
\label{table_1}
\begin{tabular}{lccc}
\hline

Parameters     & Cycle 1 & Cycle 2 & Cycle 3 \\
\hline
T                       & 7       & 9       & 10      \\
V$_{\rm LOS}$           & 2       & 4       & 6       \\
V$_{\rm turb}$          & 1       & 3       & 4       \\
B$_{\parallel}$         & 1       & 2       & 3       \\
B$_{\perp}$             & 1       & 2       & 3       \\
$\phi$                  & 1       & 1       & 1       \\  
\hline
\end{tabular}

\end{table}

\begin{figure*}[]
    \centering
    \includegraphics[width=1\linewidth]{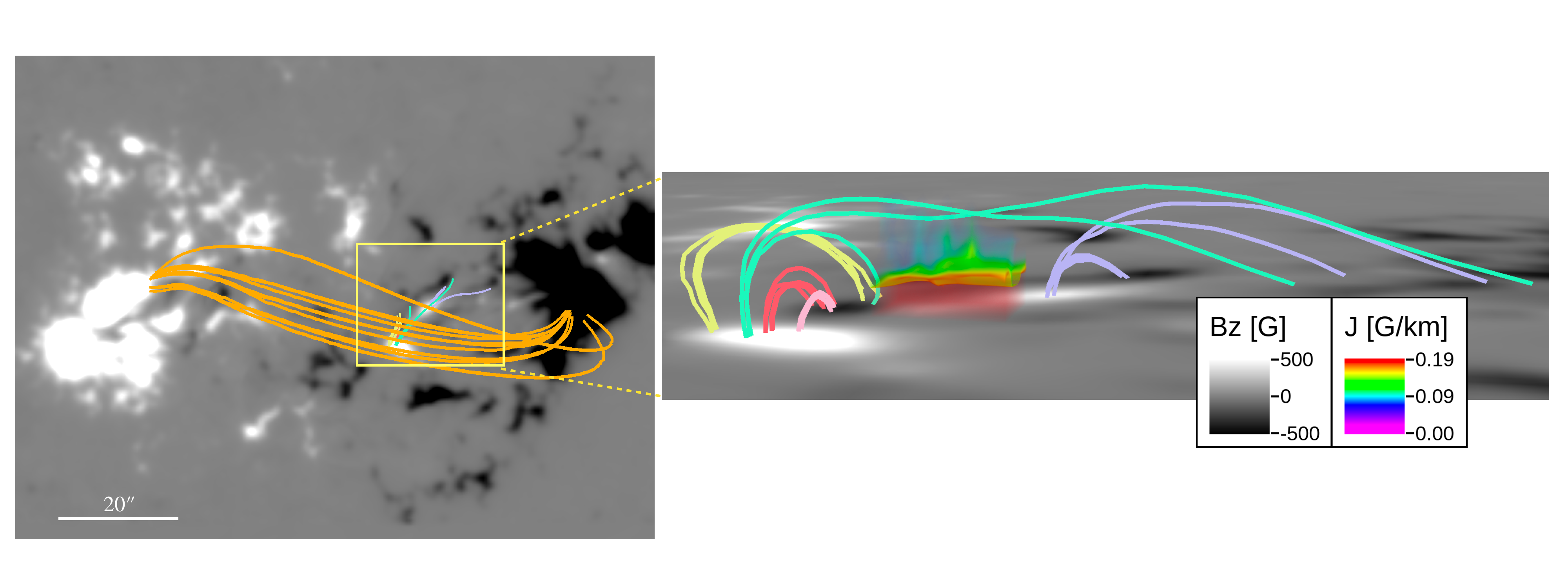}
    \caption{The magnetic field topology over the EFR.\textit{ Left panel}: The simulated field lines connecting two main opposite polarities. Small-scale loops connecting bipolar regions are highlighted by a yellow box.\textit{ Right panel:} Side view of magnetic field topology over the bipolar regions highlighted by yellow box in the right panel. For better visualization different field lines are displayed in distinct colors. The background image refer to the vertical component of magnetic field ($B_z$). The appearance of maximum current density ($J$) is also highlighted near the PIL.}
    \label{fig:cgem_simulation}
\end{figure*}

As discussed in the section, \ref{sec:method_and_data}, we inferred the physical parameters, such as temperature, magnetic field, LOS velocity, and microturbulent velocity, by inverting multiple lines (\ion{Fe}{1}~6301~\AA\ line pair, \ion{Ca}{2}~K, and \ion{Ca}{2}~8542\AA) simultaneously using the STiC code. 
We obtained the stratification of the parameters as a function of the logarithm of the optical depth scale at 500~nm (log$\tau_{500}$). 
We inverted all pixels along the selected slits passing through bipolar regions, the location of these slits can be seen in Figure \ref{fig:blos_photo_chromo_slice}. The inversion of pixels on the slits were performed using three different cycles (see Table \ref{table_1}). We note that the uncertainties in the parameters increase in higher layers ( log$\tau_{500} < -4.5$) as the observed chromospheric lines (\ion{Ca}{2}~K, and \ion{Ca}{2}~8542\AA) are not sensitive to those regions \citep{2018A&A...612A..28L, 2021A&A...649A.106Y}.
As an example the observed and best fit of the selected pixels marked in Figure~\ref{fig:invertedmaps_L1} are demonstrated in the appendix (Figure~\ref{fig:example_fittingL1}).

The stratification and temporal evolution of physical parameters, such as temperature, LOS velocity, LOS magnetic field, and microturbulent velocity, obtained along the slits at selected optical depths are shown in Figure~\ref{fig:invertedmaps_L1} (for L2--L4 slits see Figures \ref{fig:example_fittingL2}--\ref{fig:example_fittingL4} in the appendix). Notably, there is a remarkable similarity between the LOS magnetic field obtained from the non-LTE inversion and the WFA both in the photosphere and the chromosphere. In all slits, the LOS magnetic field decreases as a function of optical depth, which is expected as gas pressure decreases several orders of magnitude from the photosphere to the chromosphere, whereas magnetic pressure decays more slowly \citep{2014A&ARv..22...78W}. Therefore, the magnetic field tends to become more room-filling and weaker in the chromosphere. Additionally, the stratification in temperature exhibits an enhancement in the deeper layers (noticed log$\tau_{500} < -2$). At the log$\tau_{500} = -3$, the temperature reaches up to 8~kK. These temperature enhancements are located close to the mixed polarity regions.
Moreover, at the heated locations strong gradient in the LOS velocity is clearly visible. Moreover, the microturbulent velocity, ranges from 0 to 10~km~s$^{-1}$, normally shows enhancement at the location of strong velocity gradients and heated layers. The simultaneous presence of temperature enhancement and strong upflows and downflows suggests that magnetic energy is being released in the lower atmosphere due to magnetic reconnection, such scenario is also noticed in simulation at the site of reconnection \citep{2019A&A...626A..33H, 2022ApJ...929..103T}. 

\begin{figure*}[!ht]
    \centering
     \includegraphics[width=0.34\linewidth]{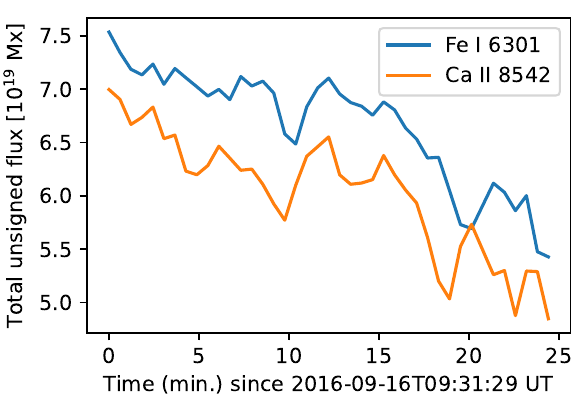}
     \includegraphics[width=0.64\linewidth]{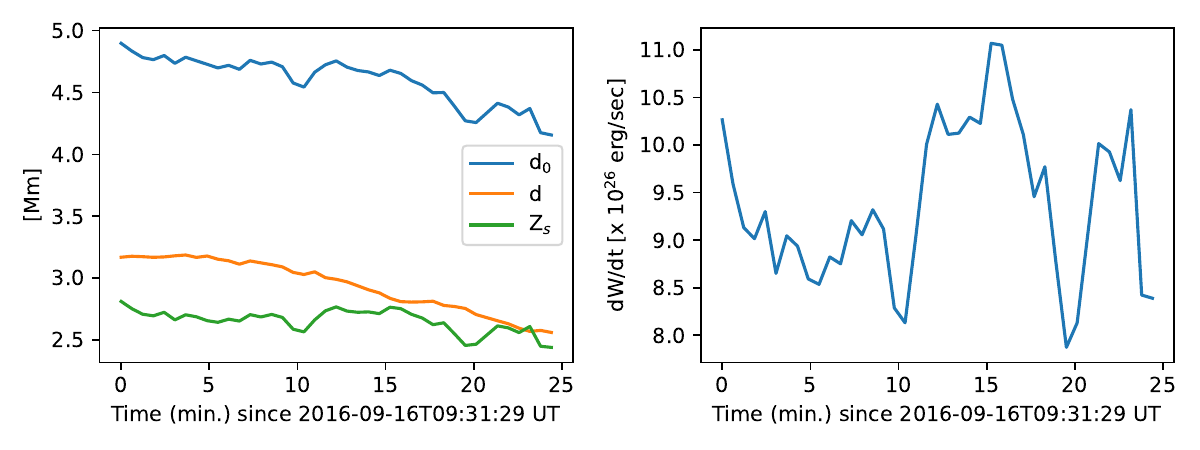}
    \caption{Temporal evolution of physical quantities for a bipolar region shown in Fig~\ref{fig:bipolar2}. \textit{Left panel:} The unsigned magnetic flux in the photosphere and chromosphere is represented by a solid blue and orange line, respectively. \textit{Middle panel:} The temporal evolution of interaction distance ($d_0$), separation between two opposite polarity (d), and the height of magnetic reconnection separator ($Z_s$). \textit{Right panel}: The total rate of magnetic field energy releases as heat, $dW/dt$. }
    \label{fig:box1_heatrate}
\end{figure*}

\subsection{Magnetic Field Topology Above Bipolar Regions}
To investigate the topology of the magnetic field above the bipolar regions, we reconstructed the 3D magnetic field lines as described in Section~\ref{sec:cgem_intro}. While the SST observed a negative polarity patch of the EFR, we considered both polarity patches from HMI magnetograms to ensure the simulation satisfied the flux-balance condition. We analyzed the magnetic field topology over a specific region (including L1 and L2 slits) that exhibited strong intensity in the chromospheric lines. This region is clearly seen in both SST/CRISP and SDO/HMI magnetograms. We note that the SST observations have better spatial resolution compared to HMI, allowing us to clearly identify small-scale bipolar regions, which are not clearly visible in the HMI magnetograms. 

As an example the obtained field configuration in the EFR is demonstrated in Figure~\ref{fig:cgem_simulation}. For visualization purpose, different field lines are highlighted by different colors. It shows that longer loops connect strong polarity locations as indicated by the orange lines in the left panel. It also shows that loops with shorter heights ($< 3$~Mm) are located above a mixed polarity regions that consists of two bipolar regions. These opposite polarity patches are also associated with the serpentine field lines, normally observed in an EFR \citep{2004ApJ...614.1099P,2018ApJ...854..174T, 2019A&A...632A.112Y}. 
The simulation illustrates that the magnetic field configuration are complex, where field lines originating from one bipolar region connect to the nearby regions.

We also observed a strong presence of currents ($J = \nabla\times B$), located ($<$~3~Mm) near the PIL of the bipolar region (see Figure~\ref{fig:cgem_simulation}). These locations of intense current are typically considered as possible locations for magnetic field reconfiguration or magnetic reconnection. The region displaying strong currents also exhibits significant intensity enhancement in the chromospheric spectral lines and AIA images (see Figure~\ref{fig:overview}). Furthermore, the presence of strong upflows and downflows, as inferred from the non-LTE inversion, suggests that magnetic reconnection likely occurred at this location. Additionally, the presence of dips in certain field lines (cyan colored lines) indicates that they have formed or rearranged after a magnetic reconnection event. This configuration bears similarity to the heating model based on reconnection proposed by \cite{2018ApJ...862L..24P}, which is discussed in the following section \ref{sec:heating_model_priest}. 

\begin{figure*}
    \centering
    \includegraphics[width=0.5\linewidth]{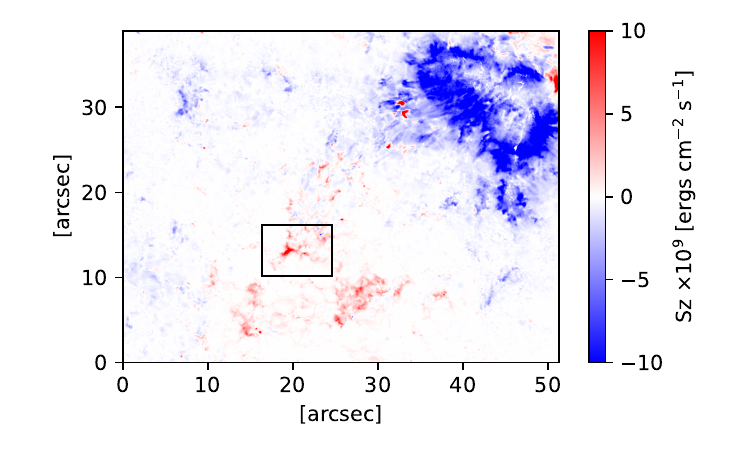}
     \includegraphics[clip,trim=0.45cm  -0.1cm 0.3cm 0cm,width=0.36\linewidth]{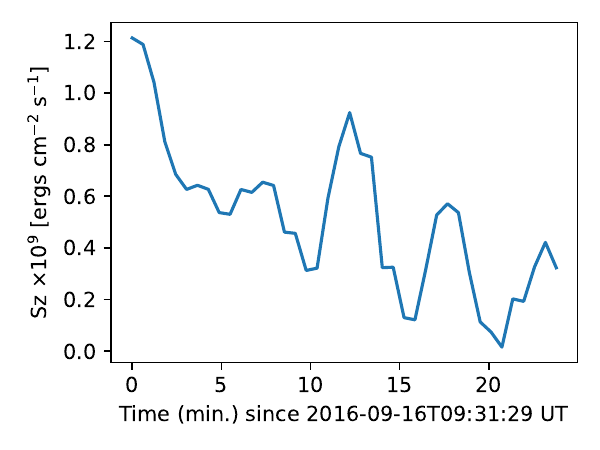}
    \caption{The vertical Poynting flux ($S_z$) in the photosphere (left panel) and the evolution of average $S_z$ (right panel) in a black box shown in the left panel.}
    \label{fig:poynting}
\end{figure*}

\subsection{Heating Model of a Converging Bipolar Region}
\label{sec:heating_model_priest}
Recently, a series of papers proposed a theoretical model to explain chromospheric and coronal heating using magnetic reconnection driven by flux cancellation \citep{ 2018ApJ...862L..24P, 2019ApJ...872...32S, 2020ApJ...891...52S}. They demonstrated that if two opposite-polarity regions, separated by a distance $d$ and having magnetic flux $\pm F$, situated below horizontal magnetic field $B_0$, will undergo reconnection if $d$ is smaller than the flux interaction distance, $d_0$ \citep{1998ApJ...507..433L}. The flux interaction distance can be expressed as follows:
\begin{equation}
    d_0 = \sqrt{\frac{F}{\pi B_0}}. 
    \label{Eq:intedist}
\end{equation}

They also derived the location of a semicircular separator in the upper atmosphere, where the magnetic field vanishes. This location is given by the following expression:
\begin{equation}
    Z_s = \sqrt{d^{2/3} d_0^{4/3} - d^{2}}.
    \label{Eq:separator}
\end{equation}

Furthermore, in the case of magnetic reconnection, the total rate of magnetic energy released as heat is given by,
\begin{equation}
    \frac{dW}{dt} = 0.4~S_i =  0.8 \frac{2\pi}{3} \frac{\nu_0 B_0^2}{\mu}d_0^2 \frac{M_{A0}}{\alpha} \frac{[1 - (d/d_0)^{4/3}]}{(d/d_0)^{2/3}},
    \label{Eq:heatrate}
\end{equation}
where $S_i$ is the Poynting flux, $\nu_0$ is converging speed of flux at the photosphere, $M_{A0}$ and $\alpha$ are Alfv\'en Mach number and constant. The derivation of above equations are given in \citealt{2018ApJ...862L..24P}. We calculated the above equation for a bipolar region by adopting $\alpha = 0.1$ and $M_{A0} = 0.1$ \citep{2014masu.book.....P, 2018ApJ...862L..24P}. These values are also adopted by \citealt{2020ApJ...897...49P} to investigate a small-scale magnetic flux cancellation event in a quiet-Sun region.

The physical quantities such as $F$, $d$, $v_0$, and $B_0$ can be derived from observations. As an example, for a bipolar patch shown in Figure~\ref{fig:bipolar2}, we estimate $F$ as half of the total unsigned magnetic flux of the patch in the photosphere. We note that the theoretical model considers equal negative and positive flux, in our case the flux is different in opposite polarity patches. To determine $d$, we use the magnetic flux-weighted centroid position of the opposite polarity (similar approach adopted by \citealt{2023ApJ...944..215Y}). Furthermore, the converging speed is estimated from the temporal evolution of $d$. To determine $B_0$, we utilize the 3D magnetic field obtained from CGEM modeling. We take the average of the horizontal magnetic field, above the selected FOV, within the height range of 10 to 15 Mm, which yield a value of 50~G. Subsequently, with these obtained values we can estimate Eqs.~\ref{Eq:intedist}--\ref{Eq:heatrate}. 

The temporal evolution of the parameters obtained for the selected patch (see Figure~\ref{fig:bipolar2}) is shown in \ref{fig:box1_heatrate}. The figure demonstrates that the magnetic flux decreases ($\sim10^{16}$~Mx~s$^{-1}$) as a function of time due to flux cancellation. Consequently, the separation between opposite-polarity patches also decreases with a converging speed of 0.4~Km~s$^{-1}$. During the SST observations, $d_0$ for this particular region stays around 4.5~Mm. We also note that $d$ value is always less than the interaction distance. This implies that this region can have magnetic reconnection in the atmosphere. The estimated height of magnetic reconnection separator, $Z_s$, is between 2 to 3~Mm, which is also in agreement with the location of strong vertical current obtained from the CGEM simulation. 

The magnetic energy released as heat during reconnection, $dW/dt$, varies from 8 to 11 $\times$ 10$^{26}$ ergs s$^{-1}$ or $\sim$3$\times$ 10$^{9}$ ergs cm$^{-2}$ s$^{-1}$ within the selected FOV. 
This value is two orders of magnitude larger than the thermal energy estimated from the DEM approach, which only considers coronal losses. Typically, the chromospheric energy losses tend to exceed coronal losses significantly \citep{1977ARA&A..15..363W}.

\subsection{The Photospheric Poynting flux}
To estimate the upward transport of magnetic energy, we estimated the vertical Poynting flux as \citep{2015PASJ...67...18W},
\begin{equation}
S_z = [v_z B_h^2 - (\mathbf{v_h} \cdot \mathbf{B_h}) B_z] 4\pi,
\label{Eq:poynting_flux}
\end{equation}

where $v_z$ and $v_h$ are the vertical and horizontal component of the velocity. $B_h$ and $B_z$ are the horizontal and vertical component of magnetic field. We obtain the vertical velocity, $v_z$, and the magnetic field vector, ($B_x$, $B_y$, $B_z$), by inverting the \ion{Fe}{1} 6301 Å line pair, obtained from the SST/CRISP, using ME approach in the photosphere (see Sect.~\ref{sec:method_and_data}). The horizontal components of velocity in the photosphere, $v_x$ and $v_y$, are determined using the FLCT method \citep{2008ASPC..383..373F} with $Bz$ component of magnetic field obtained from the SST/CRISP. 

Figure~\ref{fig:poynting} shows the estimated $S_z$ over the observed FOV. The black box shown in the figure indicates the location where intense brightening is observed in the chromospheric lines and coronal images. The temporal evolution of $S_z$ in this region demonstrates that the positive value varies around $\sim$5$\times$10$^8$ ergs cm$^2$ s$^{-1}$, which is sufficient to heat the local chromosphere and corona \citep{1977ARA&A..15..363W}. We also note that the estimated energy release from the theoretically derived equation (Eg.~\ref{Eq:heatrate}) is $\sim3\times10^{9}$~ergs cm$^2$ s$^{-1}$, whereas the Poynting flux (Eq.~\ref{Eq:poynting_flux}) is around $\sim5\times10^8$~ergs cm$^2$ s$^{-1}$ (see Fig~\ref{fig:poynting}), which is lower by a factor of six. Moreover, the fluctuations in the temporal behavior of $S_z$ demonstrate oscillatory patterns that could potentially be attributed to the five-minute oscillations occurring in the photosphere \citep{1970ApJ...162..993U}.

\section{Discussion} 
\label{sec:Discussion}
In our study, we investigated the thermal, kinematic, and magnetic structures of small-scale bipolar regions present in the vicinity of an EFR, and their impact on chromospheric and coronal heating. To achieve this, we utilized multi-line spectropolarimetric observations of an EFR located at the disk center. The observations were performed simultaneously in the \ion{Ca}{2} 8542~\AA, \ion{Ca}{2}~K, and \ion{Fe}{1}~6301~\AA\ lines using the CRISP and CHROMIS instruments at the SST.
By combining these high-resolution, multi-line observations, we infer the stratification of physical parameters such as temperature, magnetic field, LOS velocity, and micro-turbulent velocity across selected regions using the STiC, a non-LTE multi-line inversion code.

Additionally, we used co-aligned AIA images to understand the thermal distribution in the corona above bipolar events using DEM approach. Furthermore, we performed a data-driven magneto-frictional simulation to understand the magnetic field topology of these events. We also investigated how the total rate of magnetic energy released via magnetic reconnection resulting from flux cancellation contributes to chromospheric and coronal heating using cancellation nanoflare model.

Our observations demonstrated that the temporal evolution of a converging bipolar event, leading to flux cancellation at the rate of $\sim$ $10^{16}$ Mx~s$^{-1}$, not only resulted in significant brightening in the chromospheric lines (such as \ion{Ca}{2} 8542 Å and \ion{Ca}{2} K), but also in the transition region and coronal images observed by AIA. Magnetic reconnection, driven by the flux cancellation, produced detectable signatures in the chromospheric spectral lines, which exhibited complex asymmetric shapes attributed to intense heating and velocity gradients. These shapes can be attributed to the occurrence of magnetic reconnections within the lower solar atmosphere. The non-LTE inverted model atmosphere provided clear evidence of heating and strong upflows/downflows at various layers above bipolar regions.

For the selected pixels passing through bipolar regions, the stratification and temporal evolution of temperature show that the temperature in the lower chromosphere (e.g., log$\tau_{500} \sim -2$; in case of slit L2 shown in Figure~\ref{fig:invertedmaps_L1}) raised up to $\sim$8~kK. Such temperature enhancement are also observed in flares \citep{2017ApJ...846....9K,2021A&A...649A.106Y}. The location of heating does not always lie on the polarity inversion line (PIL) but is situated close to it. This could be attributed to the 3D nature of reconnection or the presence of serpentine structures in the magnetic field lines within the EFR. Furthermore, at these locations the LOS velocity exhibit both upflows (blueshift) and downflows (redshift). Typically, the upflows ($\sim$10~km~s$^{-1}$) are observed in the higher layers whereas the downflows ($\sim$10~km~s$^{-1}$) are seen in the deeper layers. Such flows can be clearly identified in the spectral profiles of selected pixels demonstrated in Figure~\ref{fig:example_fittingL1}.
 In some cases, the upflows/downflows can reach upto $\sim$20~km~s$^{-1}$ (see profiles shown in Figures~\ref{fig:example_fittingL1}-\ref{fig:example_fittingL4}). The presence of increased temperature and strong bidirectional flows provides evidence of a reconnection event in the lower solar atmosphere.

We compared our observations  with a cancellation nanoflare model given by \citealt{2018ApJ...862L..24P}. According to this model, two converging bipolar regions on the photosphere will reconnect, if their separation is below the interaction distance ($d_0$), and thus can heat the upper atmosphere. We observationally derived $d_0$, magnetic reconnection separator ($Z_s$), and the rate of magnetic field energy released as heat ($dW/dt$). In our selected bipolar region, we find that their separation is always less than $d_0$, and the field lines associated with them are likely to reconnect somewhere in the chromosphere. The estimated, $Z_s$ value that tells the location of magnetic reconnection is around 2.5-3~Mm. This height range is also in agreement with the location of total vertical currents derived from the data-driven simulation.

The estimated released energy $dW/dt$ in the selected region is approximately $\sim$3$\times$ 10$^{9}$ ergs~cm$^{-2}$~s$^{-1}$. This value is roughly six times more than the energy flow estimated using the Poynting flux (see Eq. \ref{Eq:poynting_flux}). This difference could be due to the simplistic assumption of a bipolar structure consisting of two opposite polarity sources with equal flux in the theoretical derivation. However, in reality, the photospheric structure is much more complex. Our observations demonstrate the presence of two bipolar regions with different flux in each polarity. Unlike the theoretical model proposed by \citealt{2018ApJ...862L..24P}, the field lines obtained from data-driven simulations reveal connections to nearby opposite field locations. The theoretical heating model based on reconnection may need to incorporate multiple bipolar structures to accurately simulate real observations and explain the various aspects of heating caused by reconnection in the lower solar atmosphere. 

 \raggedbottom
\section{Conclusion}
\label{sec:conclusion}
Our study provides valuable insights into the thermal, kinematic, and magnetic structures of small-scale bipolar regions in a flux emerging region and their influence on chromospheric and coronal heating. Utilizing multi-line spectropolarimetric observations, non-LTE inversions, co-aligned AIA images, and data-driven magneto-frictional simulations, we found that converging bipolar events, resulting in flux cancellation and magnetic reconnection, led to significant chromospheric brightening and complex spectral signatures. This resulted in significant heating and velocity gradients in the lower solar atmosphere. The observed temperature enhancement, upflows, and downflows retrieved from non-LTE inversion also suggest that the magnetic reconnection is occuring in the lower solar atmosphere, specially around the
temperature minimum and above it. The released magnetic energy from flux cancellation was shown to sufficiently heat the local chromosphere and corona. 

We note that the spectropolarimetric signals in the \ion{Ca}{2} 8542~\AA\ line were inadequate for deriving the horizontal magnetic field in the chromosphere. To fully understand the magnetic field changes associated with magnetic reconnection and heating mechanisms, it is critical to obtain sufficient linear polarization signals in both the photosphere and chromosphere. State-of-the-art instruments and new generation solar telescopes like the Daniel K. Inouye Solar Telescope (DKIST; \citealt{2015csss...18..933T}) and the European Solar Telescope (EST; \citealt{2016SPIE.9908E..09M}) will be instrumental in achieving this goal.

\begin{acknowledgements}
We would like to thank the anonymous referee for the comments and suggestions. The Swedish 1-m Solar Telescope is operated on the island of La Palma by the Institute for Solar Physics of Stockholm University in the Spanish Observatorio del Roque de los Muchachos of the Instituto de Astrof\'isica de Canarias. The Institute for Solar Physics is supported by a grant for research infrastructures of national importance from the Swedish Research Council (registration number 2021-00169).
We acknowledge support from NASA LWS NNH17ZDA001N, NASA LWS 80NSSC19K0070, NASA ECIP 80NSSC19K0910, NASA HSR NNH21ZDA001 and NSF CAREER award SPVKK1RC2MZ3 (R.Y. and M.D.K.). Resources supporting this work were provided by the NASA HighEnd Computing (HEC) Program through the NASA Advanced Supercomputing (NAS) Division at Ames Research Center. 
This work utilized the Alpine high performance computing resource at the University of Colorado Boulder. Alpine is jointly funded by the University of Colorado Boulder, the University of Colorado Anschutz, Colorado State University, and the National Science Foundation (award 2201538).
This project has received funding from the European Research Council (ERC) under the European Union's Horizon 2020 research and innovation program (SUNMAG, grant agreement 759548). 
We acknowledge the use of the visualization software VAPOR \citep{2019Atmos..10..488L} for generating relevant graphics. Data and images are courtesy of NASA/SDO and the HMI and AIA science teams. 
This research has made use of NASA’s Astrophysics Data System. We acknowledge the community effort devoted to the development of the following open-source packages that were used in this work: NumPy (\url{numpy.org}), matplotlib (\url{matplotlib.org}) and SunPy (\url{sunpy.org}).

\end{acknowledgements}

\bibliographystyle{aa}
\bibliography{new-ref}  

\appendix 

\section{Fitting of observed profiles and retrieved model atmosphere}
\begin{figure}[!h]
    \centering
    \includegraphics[width=1\linewidth]{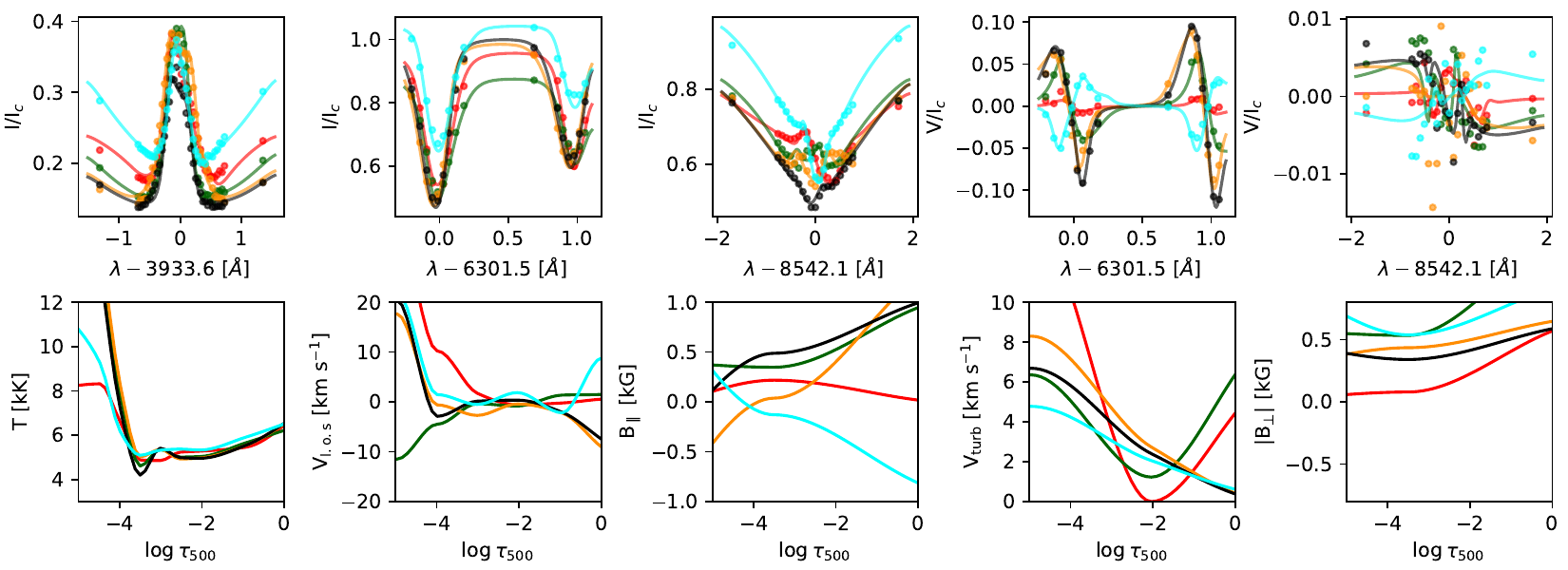}
    \caption{Fitting of observed Stokes profiles and retrieved model atmosphere. \textit{Top panel}: Observed (dotted lines) and best fit (solid colored lines) profiles of pixels located at different positions of L1 slit, where the color correspond to the colored cross symbols marked in Figure~\ref{fig:invertedmaps_L1}. \textit{Bottom panels}: The retrieved stratification of physical parameters obtained after fitting observed Stokes profile.}
    \label{fig:example_fittingL1}
\end{figure}

Figure \ref{fig:example_fittingL1} shows the observed and best-fit profiles of the selected pixels using STiC code, which are indicated by cross symbols in Figure \ref{fig:invertedmaps_L1}. Each pixel's profile is represented by a different color that matches the corresponding marked pixel.

\begin{figure}
    \centering
    \includegraphics[width=1\linewidth]{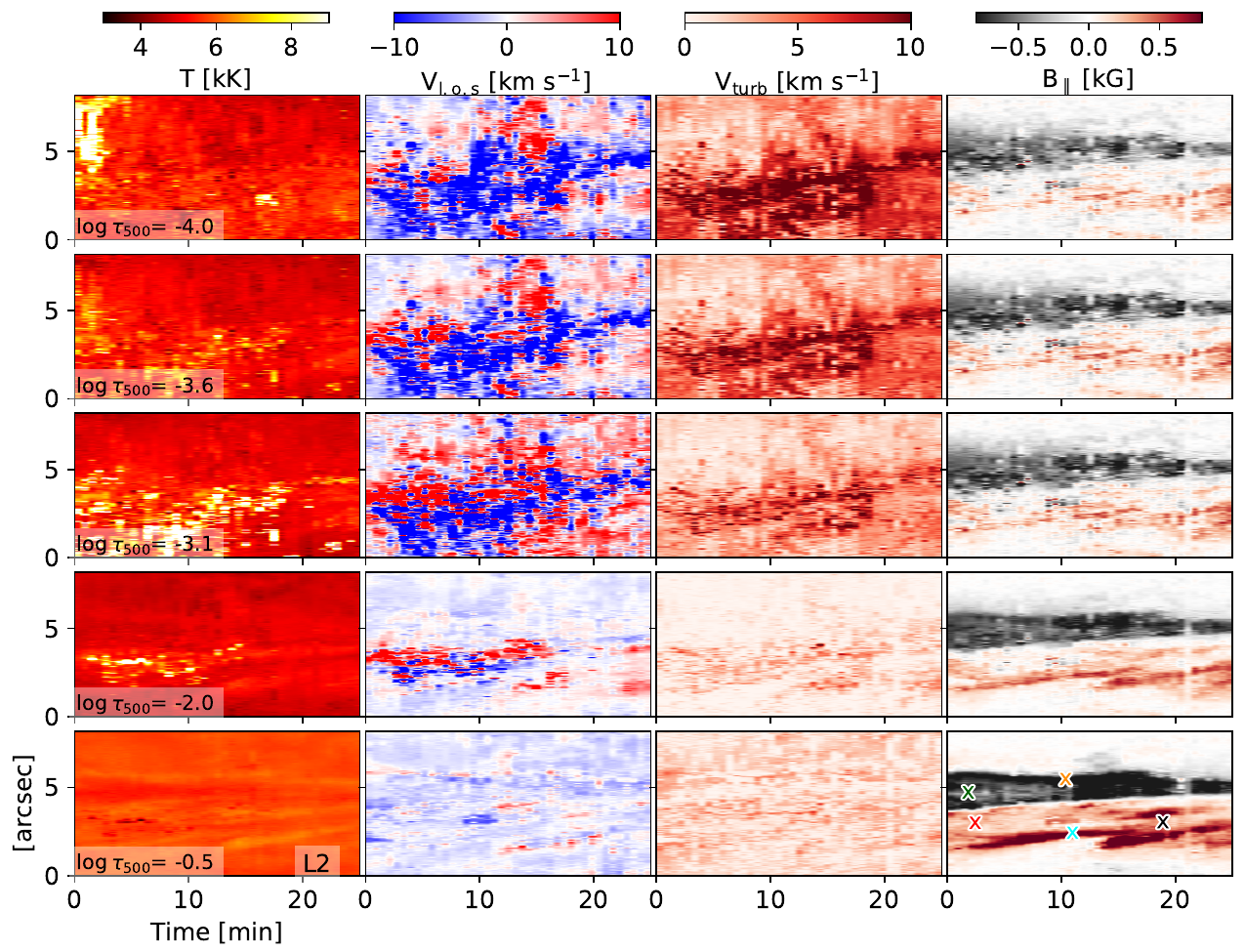}
    \includegraphics[width=1\linewidth]{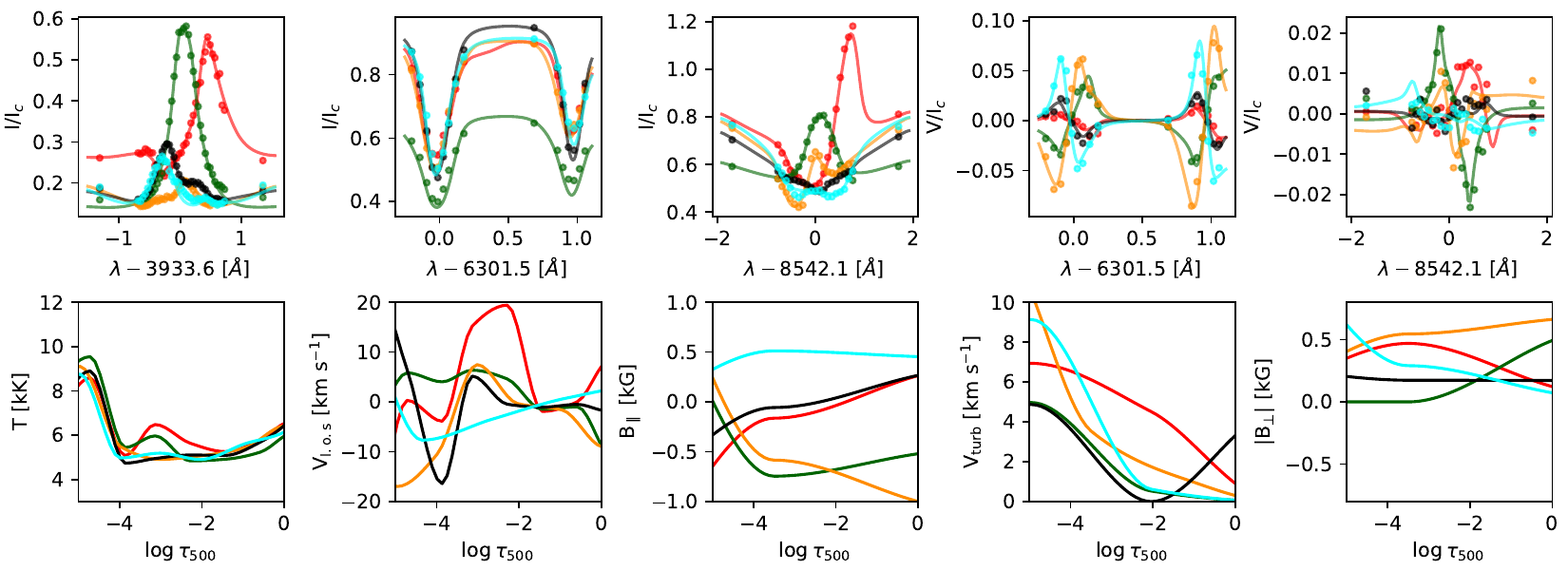}
    \caption{Temporal evolution and stratification of physical parameters obtained from non-LTE inversion for the L2 slit highlighted in Figure~\ref{fig:blos_photo_chromo_slice}. The fitting of observed Stokes profiles and retrieved parameters at the location of colored cross symbols are shown in the bottom panels.}
    \label{fig:example_fittingL2}
\end{figure}

\begin{figure}
    \centering
    \includegraphics[width=1\linewidth]{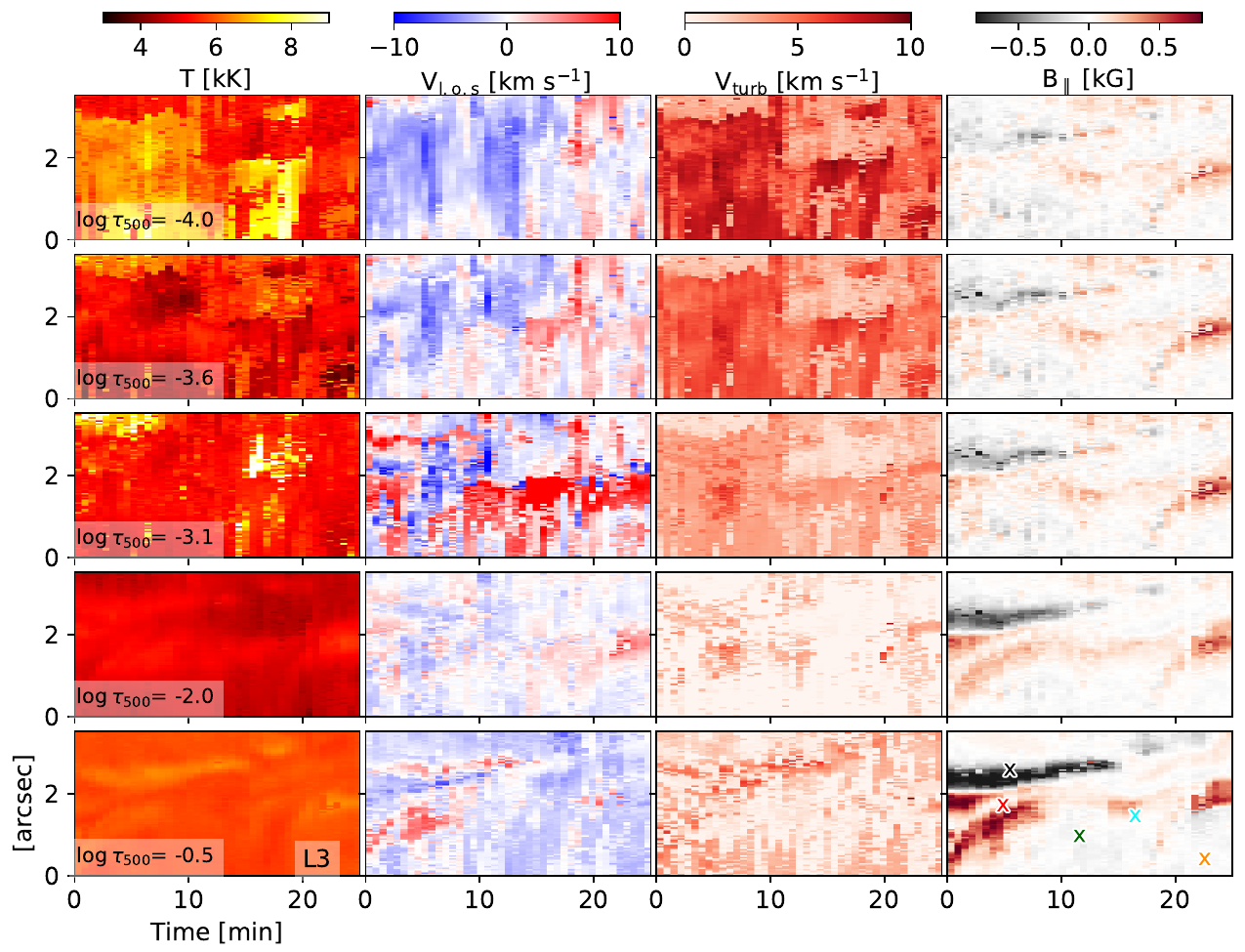}
    \includegraphics[width=1\linewidth]{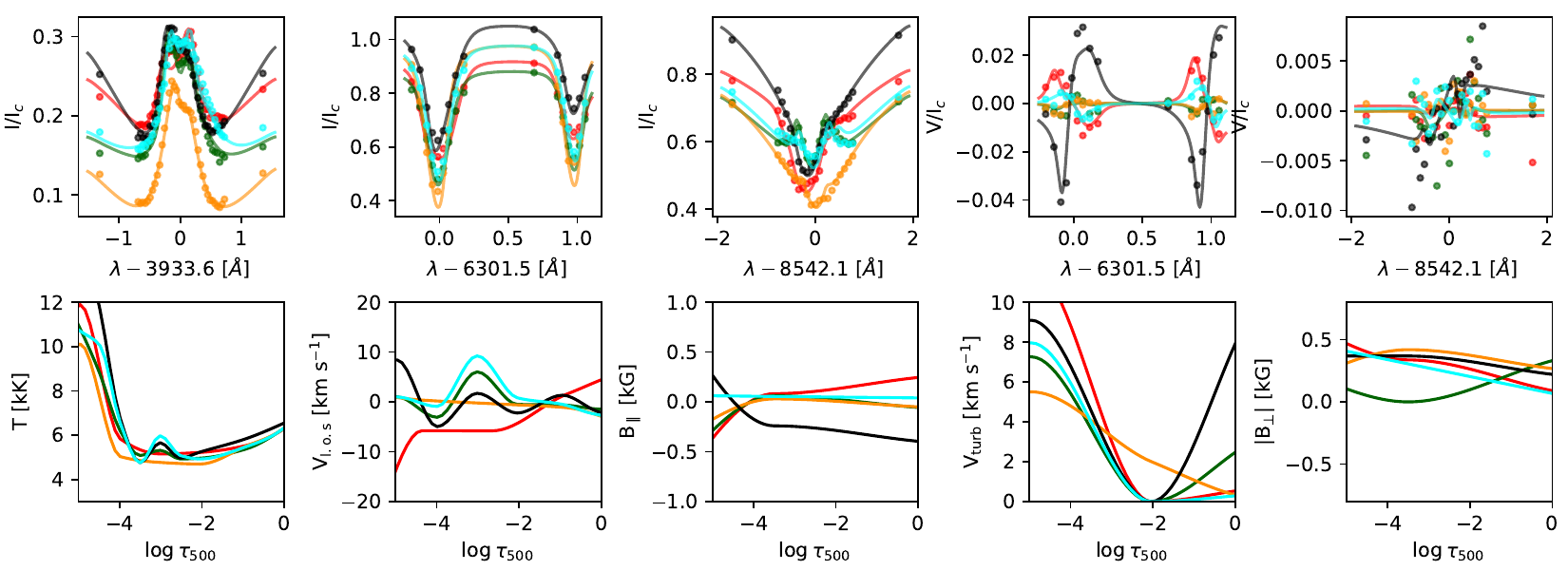}
    \caption{Same as Figure~\ref{fig:example_fittingL2}, but for L3 slit.}
    \label{fig:example_fittingL3}
\end{figure}

\begin{figure}
    \centering
    \includegraphics[width=1\linewidth]{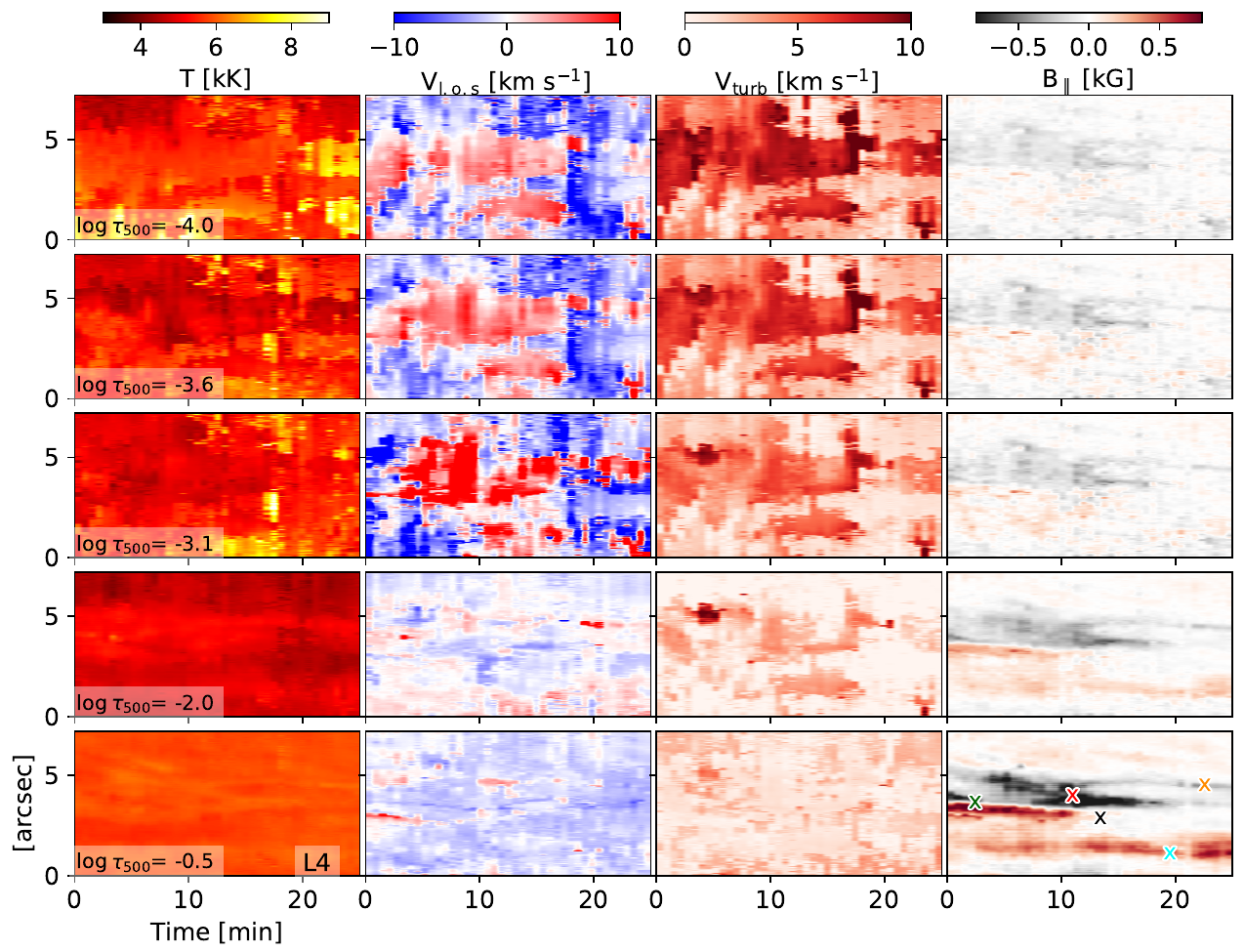}
    \includegraphics[width=1\linewidth]{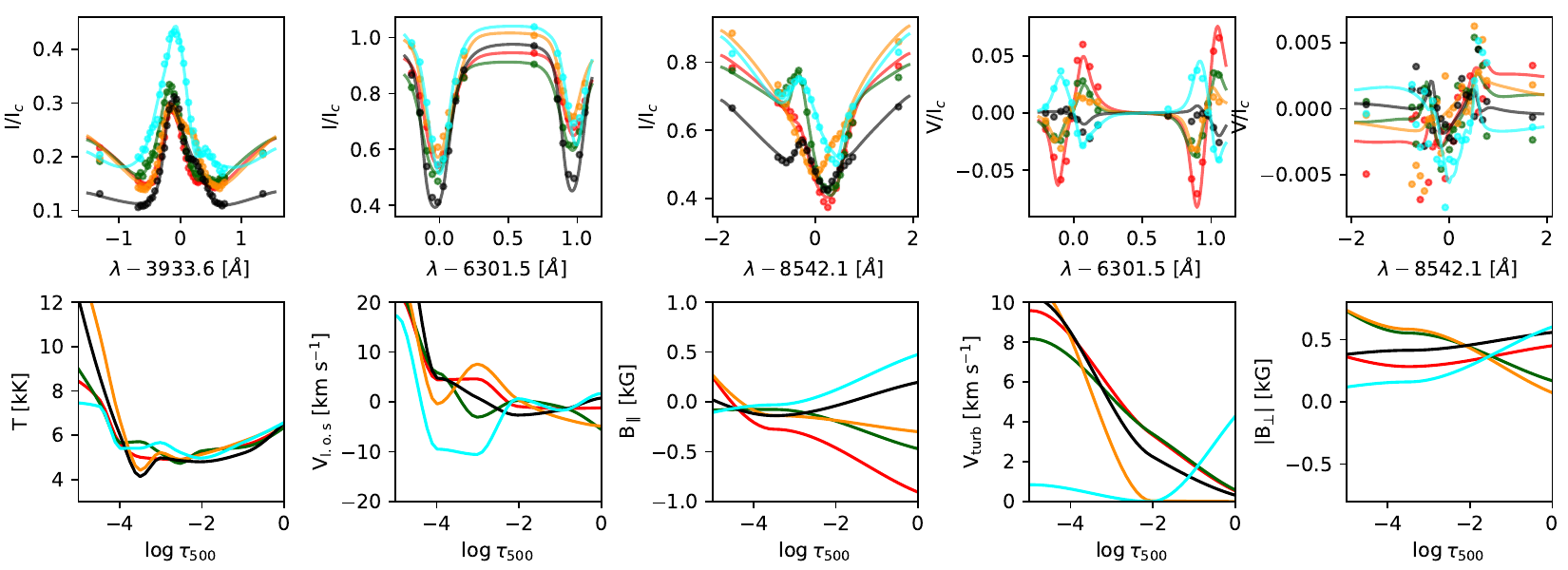}
    \caption{Same as Figure~\ref{fig:example_fittingL2}, but for L4 slit.}
    \label{fig:example_fittingL4}
\end{figure}

Figures \ref{fig:example_fittingL2} -- \ref{fig:example_fittingL4} depict the stratification and temporal evolution in physical parameters, such as temperature, LOS velocity, LOS magnetic field, and microturbulent velocity. These parameters are derived along the slits (L2, L3, and L4) highlighted in Figure~\ref{fig:blos_photo_chromo_slice}. Additionally, the observed and best fitted profiles, along with the retrieved model atmosphere, are demonstrated as examples for selected pixels (indicated by cross symbols).

\end{document}